%% file: 2020_BendingBehaviorOfAMLatticeStructures.tex
\documentclass[
               DIV=15, font size=10.5pt,
                 onecolumn,
               ]{scrartcl}  

\input{template/packages}
\input{template/pgfplotsettings}

\input{template/commonCommands}
\input{template/titleAndAuthors}
\input{template/journalFooter}
\usepackage{ulem}
\makeatletter
\@addtoreset{footnote}{section}
\makeatother
\crefname{figure}{Fig.}{Fig.}
\crefname{equation}{Eq.}{Eq.}
\crefname{table}{Tab.}{Tab.}

\drawboundingboxtrue
\drawboundingboxfalse 

\newcommand*{\figref}[2][]{%
	\hyperref[{fig:#2}]{%
		Fig.~\ref*{fig:#2}%
		\ifx\\#1\\%
		\else
		\,#1%
		\fi
	}%
}
\definecolor{changes}{RGB}{0,0,0}
\definecolor{changez}{RGB}{0,0,0}

\begin{document}  

\normalem
\maketitle  
  
\vspace{-1.5cm} 
\hrule 
\input{template/abstract}
 \vspace{.2cm} 
\vspace{0.25cm}
\noindent \textit{Keywords:} \input{template/keywords} 
\vspace{0.35cm}
\hrule 
\vspace{0.15cm}
\captionsetup[figure]{labelfont={bf},name={Fig.},labelsep=colon}
\captionsetup[table]{labelfont={bf},name={Tab.},labelsep=colon}
\tableofcontents
\vspace{0.5cm}
\hrule 
	\input{sections/introduction/introduction}
	\input{sections/numerics/numerics}
	\input{sections/experiments/experiments}
	\input{sections/results/results}
	\input{sections/conclusion/conclusion}

\section*{Acknowledgements} 
\input{template/acknowledgements}
\section*{Data Availability}
{
	The raw and the processed data required to reproduce these findings cannot be shared at this time due to technical limitations. Furthermore, the data also forms part of an ongoing study.
}
\newpage
\bibliographystyle{apalike}
 \bibliography{library}

\end{document}

%% file: template/packages.tex
\usepackage[dvipsnames]{xcolor}
\usepackage{psfrag} 
\usepackage{subfig}
\usepackage{hyperref}
\usepackage{amsmath}
\usepackage{graphics}
\usepackage{mathtools}
 \usepackage{pstool} 
\usepackage{todonotes}
\usepackage{gensymb}
\usepackage{bm}
\usepackage{multirow}
\usepackage{cancel}
\usepackage{float}
\usepackage{textcomp}
\usepackage{xcolor}
\usepackage[square, numbers,sort&compress]{natbib}
 
\usepackage{pgfplots}
\usepackage{pgfplotstable}
\usepgfplotslibrary{groupplots}
\usepackage{filecontents}
\usepackage{algorithm2e}
\usepackage{algorithmic}
\usepackage{siunitx}
\usepackage{tikz}
\usepackage{tikzscale}
\usepackage{tabularx} 
\usetikzlibrary{spy}
\usetikzlibrary{snakes,arrows,shapes}    
\usetikzlibrary{spy}
\usetikzlibrary{backgrounds}  
\usetikzlibrary{decorations}
\usetikzlibrary{positioning}
\usetikzlibrary{patterns}
\usetikzlibrary{calc} 

\usepackage{booktabs}   
\usepackage{makecell} 
\usepackage{colortbl}   

\usepackage{xspace}   
\usepackage{setspace}   
 
\usepackage{soul}
\usepackage{ulem}
\usepackage{currfile}
\usepackage{import}

\usepackage{authoraftertitle} 
\usepackage{authblk} 
\usepackage{cleveref}

%% file: template/pgfplotsettings.tex
\pgfplotsset{compat=1.8}

\newcommand{\findmax}[3]{
    \pgfplotstablesort[sort key={#2},sort cmp={float >}]{\sorted}{#1}%
    \pgfplotstablegetelem{0}{#2}\of{\sorted}%
    \let #3=\pgfplotsretval%
}

\pgfplotscreateplotcyclelist{myCycleListBlackAndWhite}
{%
  {black, mark=o},
  {black, mark=square},
  {black, mark=triangle},
  {black, mark=diamond}, 
  {black, mark=pentagon}, 
  {black, mark=*},
  {black, mark=square*},
  {black, mark=triangle*},
  {black, mark=diamond*}, 
  {black, mark=pentagon*}, 
}

\definecolor{darkgreen}{rgb}{0,0.4,0} 
\definecolor{darkbrown}{rgb}{0.5, 0.396, 0.09}

\definecolor{c1}{rgb}{0.0, 0.4196078431372549, 0.6431372549019608}
\definecolor{c2}{rgb}{1.0, 0.5019607843137255, 0.054901960784313725}
\definecolor{c3}{rgb}{0.6705882352941176, 0.6705882352941176,
0.6705882352941176} \definecolor{c}{rgb}{0.34901960784313724, 0.34901960784313724, 0.34901960784313724}
\definecolor{c4}{rgb}{0.37254901960784315, 0.6196078431372549,
0.8196078431372549} \definecolor{c}{rgb}{0.7843137254901961, 0.3215686274509804, 0.0}
\definecolor{c5}{rgb}{0.5372549019607843, 0.5372549019607843,
0.5372549019607843} \definecolor{c}{rgb}{0.6352941176470588, 0.7843137254901961, 0.9254901960784314}
\definecolor{c6}{rgb}{1.0, 0.7372549019607844, 0.4745098039215686}
\definecolor{c7}{rgb}{0.8117647058823529, 0.8117647058823529,
0.8117647058823529}

\pgfplotscreateplotcyclelist{myCycleListColor}
{%
  {blue, mark=o},
  {red, mark=square},
  {darkbrown, mark=triangle},
  {darkgreen, mark=diamond},
  {black, mark=pentagon},
  {blue, mark=*},
  {red, mark=square*},
  {darkbrown, mark=triangle*},
  {darkgreen, mark=diamond*},
  {black, mark=pentagon*},
}

\pgfplotsset{every axis/.append style= 
              {
                font=\small,
                mark size=2,
                line width = 0.1,
                legend style={font=\small, mark size=3, draw=none, fill=none},
                legend cell align=left,
                cycle list name=myCycleListColor,
              }
            }
            
\pgfplotstableset
{
  every odd row/.style={before row={\rowcolor[gray]{0.9}}},
  every head row/.style=
  {
    before row=
    {%
      \toprule
    },
    after row=\midrule
  },
  every last row/.style=
  {
    after row=\bottomrule
  }
}

\newif\ifdrawboundingbox

\usetikzlibrary{external} 
\tikzset{external/system call={pdflatex \tikzexternalcheckshellescape
-halt-on-error -interaction=batchmode -jobname "\image" "\texsource"}} 

\pgfkeys
{ 
  /pgf/number format/.cd,
  fixed,
  set thousands separator={\,},
  1000 sep in fractionals,
}

%% file: template/commonCommands.tex
\newcolumntype{C}[1]{>{\centering\arraybackslash}m{#1}}
\newcolumntype{R}[1]{>{\raggedright\arraybackslash}m{#1}}
\newcolumntype{L}[1]{>{\raggedleft\arraybackslash}m{#1}}

\newcommand{\delete}[1]{\xspace}

\definecolor{Reviewer1}{rgb}{0.0, 0.0, 1.0}
\definecolor{Reviewer2}{rgb}{0.0, 0.5, 0.0}

%% file: template/titleAndAuthors.tex
\title{Bending behavior of additively manufactured lattice structures: numerical characterization and experimental validation}

\author[1]{N. Korshunova\thanks{\href{mailto:nina.korshunova@tum.de}{\texttt{nina.korshunova@tum.de}},
    Corresponding author}}
\author[2]{G. Alaimo}
\author[4]{S. B. Hosseini}
\author[2]{M. Carraturo}
\author[2]{A. Reali}
\author[4]{J. Niiranen}
\author[2]{F. Auricchio}
\author[3]{E. Rank}
\author[1]{S. Kollmannsberger}

 \affil[1]{Chair of Computational Modeling and Simulation,
 Technische Universit\"at M\"unchen, Germany}
 \affil[2]{Department of Civil Engineering and Architecture, University of Pavia, Italy}
 \affil[3]{Institute for Advanced Study, Technische Universit\"at M\"unchen, Germany}
 \affil[4]{Department of Civil Engineering, Aalto University, Finland}

\newcommand{\journal}{Materials \& Design}
\newcommand{\publicationDate}{\today}

%% file: template/journalFooter.tex
\usepackage{fancyhdr}
\date{}
\fancypagestyle{plain}{%
  \fancyhf{}%
  \fancyfoot[L]{
  \vspace{-1.25cm} 
  \footnotesize{Preprint submitted to \textit{\journal{}}  \hfill
  \publicationDate
  \\
  }} }

%% file: template/abstract.tex
\section*{Abstract}
{
Selective Laser Melting (SLM) technology has undergone significant development in the past years providing unique flexibility for the fabrication of complex metamaterials such as octet-truss lattices. However, the microstructure of the final parts can exhibit significant variations due to the high complexity of the manufacturing process. Consequently, the mechanical behavior of these lattices is strongly dependent on the process-induced defects, raising the importance on the incorporation of as-manufactured geometries into the computational structural analysis. This, in turn, challenges the traditional mesh-conforming methods making the computational costs prohibitively large. In the present work, an immersed image-to-analysis framework is applied to efficiently evaluate the bending behavior of AM lattices. To this end, we employ the Finite Cell Method (FCM) to perform a three-dimensional numerical analysis of the three-point bending test of a lattice structure and compare the as-designed to as-manufactured effective properties. Furthermore, we undertake a comprehensive study on the applicability of dimensionally reduced beam models to the prediction of the bending behavior of lattice beams and validate classical and strain gradient beam theories applied in combination with the FCM. The numerical findings suggest that the SLM octet-truss lattices exhibit size effects, thus, requiring a flexible framework to incorporate high-order continuum theories.

}

%% file: template/keywords.tex
additive manufacturing, metamaterials, octet-truss lattice, Finite Cell Method, computed tomography, beam theories, strain gradient elasticity, Finite Element Method

%% file: sections/introduction/introduction.tex
\section{Introduction}
{
Mechanical metamaterials have received much attention in the past decades~\cite{Rashed2016,Schaedler2016}. One of the most common examples are octet-truss lattices. These regular, periodic structures are attractive for many industries due to the possibility of largely decoupling the effective stiffness and strength from relative density~\cite{Deshpande2001, Omasta2017,Sha2018, Zheng2014}. One further advantage of the octet-truss lattices is the possibility to relate their mechanical properties to the truss topology and geometry (see e.g.~\cite{Deshpande2001, Latture2018, Omasta2017,Tancogne2018}). Although this relation facilitates their design for specific applications, some geometrical constraints push traditional manufacturing techniques of octet-truss lattices to their boundaries. 

Recent developments in additive manufacturing have provided a unique possibility to produce such metamaterials at very small scales. Yet, the design freedom comes at the cost of process complexity. The process-induced features, even defects, often occur in the produced structures, especially metal lattices, thus altering the mechanical behavior of final parts~\cite{Echeta2020, Haubrich2019, Lozanovski2019, Maconachie2019, Duplesis2020}. Therefore, to achieve a reliable prediction of the effective properties of these imperfect structures, as-manufactured geometries should be incorporated into computer-aided engineering (CAE). One of the common ways to acquire the as-manufactured AM geometry is to perform a Computed Tomography (CT) scan~\cite{Duplesis2020,Echeta2020,Yan2012}. The scanned images provide extensive information about the microstructure of 3D printed components up to a scan resolution in the order of few microns. Thus, the CT-based analysis could lead to a better prediction of the mechanical behavior of 3D printed structures. 

In the present work, we focus on the effective bending behavior of octet-truss lattices. The most common numerical approaches for its prediction are three-dimensional (3D) Finite Element Analyses (FEA) and the application of one-dimensional (1D) beam theories. These techniques represent the two engineering extremes: one provides the most realistic solution, while the other delivers a fast and quick approximation. 3D and 1D numerical analyses are commonly used in different areas of engineering. Each of them faces major challenges when applied to additively manufactured metamaterials.

To make CT images suitable for a traditional mesh-conforming three-dimensional analysis, geometry reconstruction and mesh generation are required~\cite{Liu2017, Maconachie2019, Vayssette2019}. These steps tend to become especially laborious when metamaterials are considered. Consequently, the numerical studies are often conducted only on some specific regions of the lattices, for example on the periodic representative volumes, or by modifying the idealized CAD models~\cite{Dallago2019a, Lei2019, Liu2017, Lozanovski2019}. To overcome these long and tedious steps, a class of immersed domain methods has been developed. Immersed domain methods separate the geometrical representation from the applied discretization, thus, eliminating the necessity of geometry reconstruction and simplifying the mesh generation process. In the present work, the Finite Cell Method (FCM) is employed to perform numerical analysis directly on CT scan images of as-manufactured octet-truss lattices~\cite{Duster2017, Parvizian2007}.

Concerning 1D analyses, the conventional continuum beam theories are not necessarily applicable to the evaluation of the metamaterial or effective bending behavior. They strongly rely on the assumption of the separation of scales, i.e., the microstructural characteristic length should be much smaller than the size of the representative volume element. Nevertheless, it has been determined experimentally and numerically that these components cannot be described by conventional continuum models, such as e.g. Euler-Bernoulli or Timoshenko beam theories, when the size of the periodic cell approaches the typical wavelength of the variation of the macroscopic mechanical fields. Such deviations are normally referred to as size effects. These effects can arise at different scales. When lattice or foam-like structures are considered, size effects can occur at the scale of millimeters \citep{Onck2016}. If this scale is comparable to the component dimension, size effects are crucial for the evaluation of the part behavior. In metamaterials, size effects become especially pronounced when the corresponding structures are loaded in shear or bending~\cite{Yoder2018}. As an example, when lattice beams are considered, the relative bending rigidity increases significantly when the size of the representative cell of the lattice approaches the thickness of the beam structure. This occurs if the beam structure is composed of very few layers of lattice cells in the thickness direction~\cite{Khakalo2018, Khakalo2019}. In such scenarios, the strain gradient extensions of the classical continuum models are proven to be accurate in predicting the mechanical behavior of size-dependent lattice structures. These beam theories are especially relevant when additively manufactured lattices are analyzed as the produced scales are rather small. However, as they require the effective Young's and shear moduli as input parameters, to the knowledge of the authors of this paper they have not been validated for the as-manufactured octet-truss lattices. 

With this in mind, we aim to demonstrate and experimentally validate the proposed CT-based numerical framework which allows us to accurately evaluate the bending behavior of as-manufactured octet-truss lattice structures. To this end, the framework provides an efficient tool to compare the as-designed to as-manufactured properties under loading. Additionally, we investigate and validate the accuracy of the classical and the strain gradient beam theories by comparing their bending properties to the direct 3D numerical analysis of the as-manufactured and as-designed octet-truss lattice beams.

The present article is organized as follows. In~\cref{sec:NumericsFCM} we start with a brief description of the Finite Cell Method for the numerical analysis of as-manufactured AM lattices. Then,~\cref{sec:strainGradientTheory} recalls the fundamentals of the classical and the strain gradient beam theories. In~\cref{sec:experimentalTest}, details on the manufacturing process and the experimental setup of the bending tests are given. Further, the numerical findings are discussed in~\cref{sec:numericalInvestigation}. This section starts by comparing the as-designed to as-manufactured bending behavior and validating the results of the three-dimensional numerical analysis (see~\cref{subsec:bendingTest}). The next~\cref{subsec:strainTheoryResults} validates the strain gradient beam theories and provides a discussion on their applicability to octet-truss lattices produced by additive manufacturing. Finally, our conclusions are drawn in~\cref{sec:conclusions}.

	
}

%% file: sections/numerics/numerics.tex
\section{The Finite Cell Method for numerical analysis of CT scans }
\label{sec:NumericsFCM}
{

 The main idea beneath the Finite Cell Method is illustrated in~\cref{fig:FCMIdea}.
	
	\begin{figure}[H]
		\centering
		\hspace*{-.8cm}
		\subfloat[Physical domain $\Omega$]{
			\def\svgwidth{.32\textwidth}
			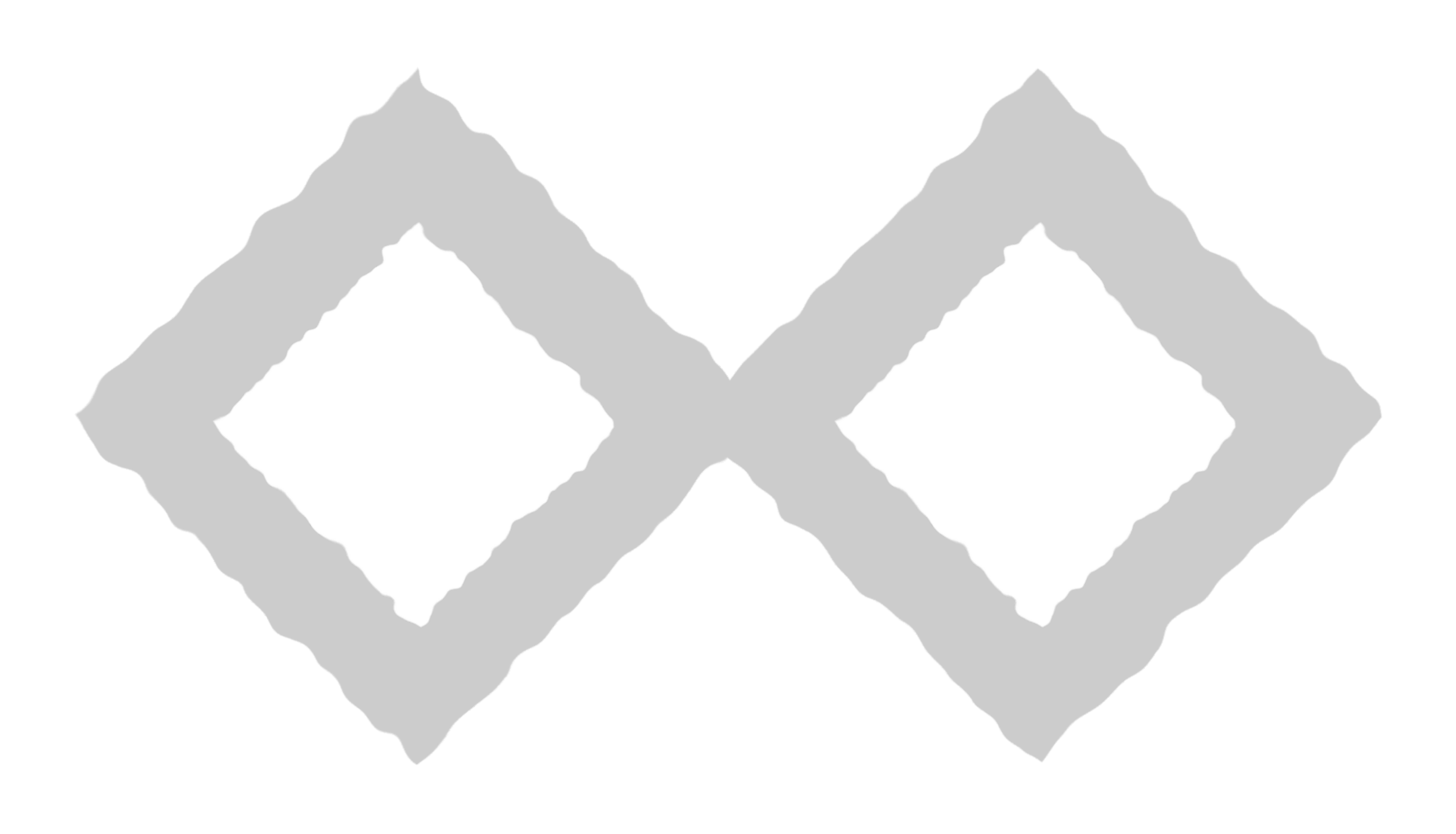
			\label{fig:PhysicalDomain} }
		\subfloat[Extended domain $\Omega_{e}$]{
			\def\svgwidth{.32\textwidth}
			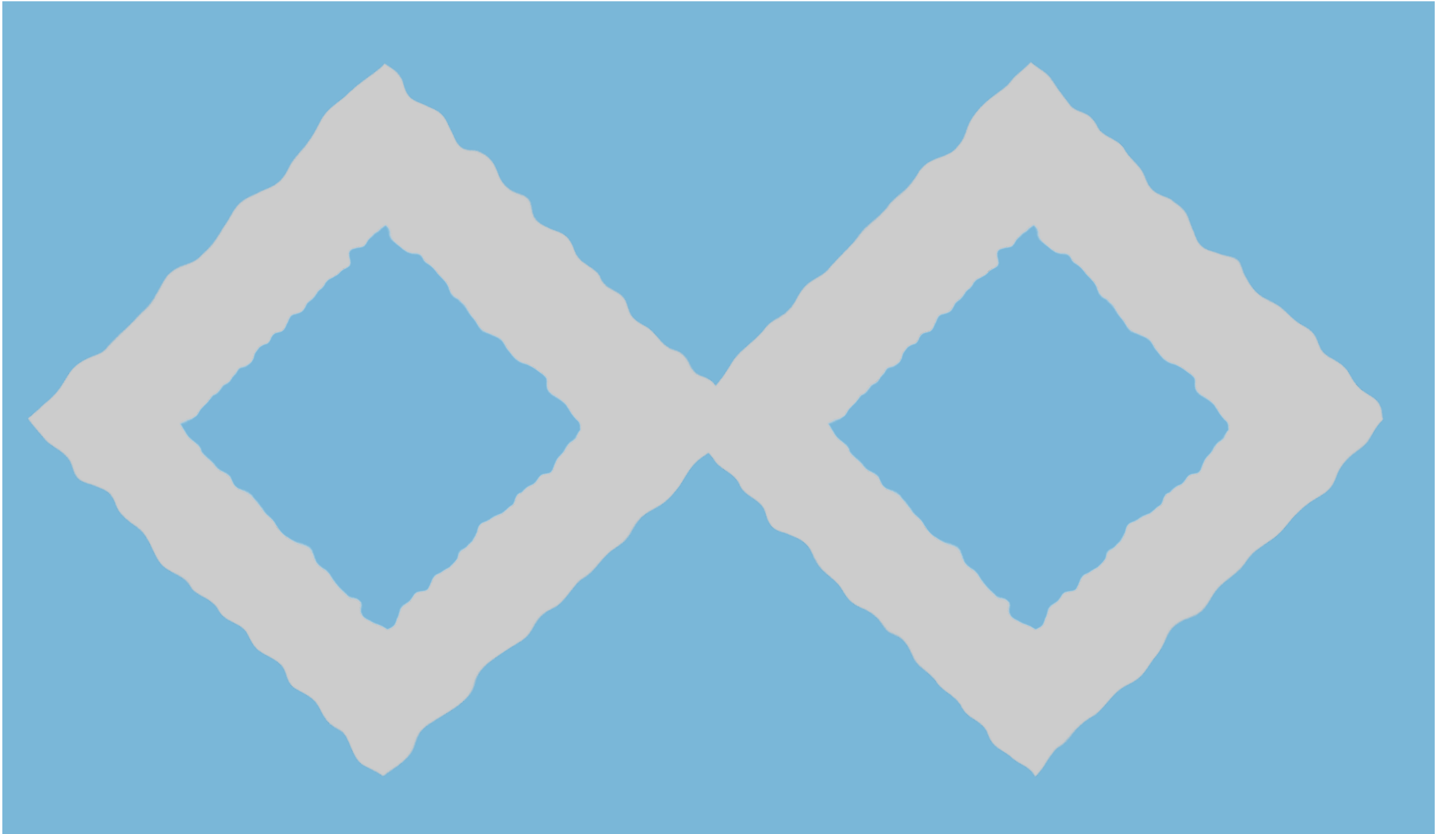
			\label{fig:ExtendedDomain} }
		\subfloat[FCM mesh (in bold) and indicator function]{
			\def\svgwidth{.32\textwidth}
			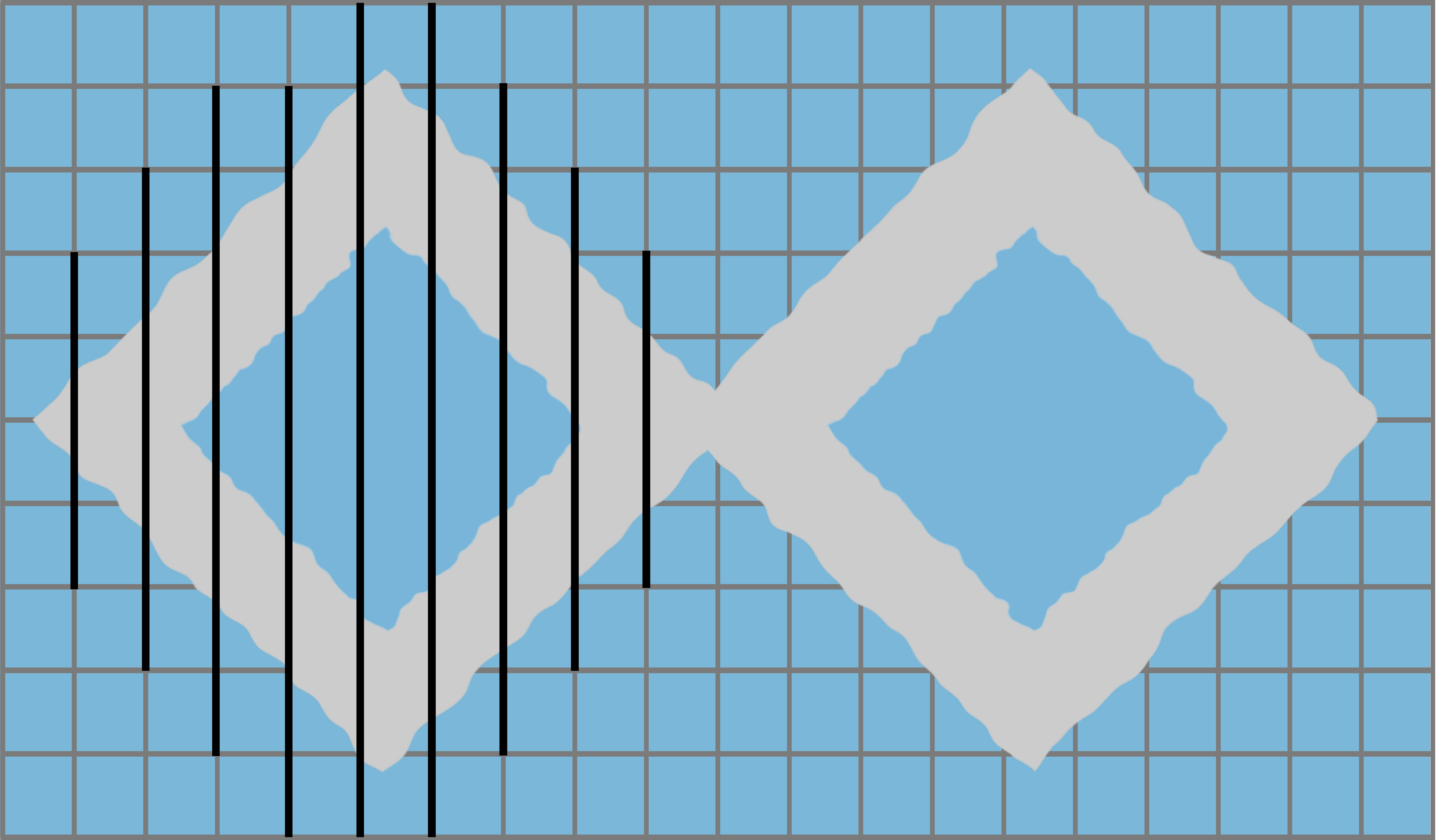
			\label{fig:FCMMeshIndicator}
		}
		\caption{The idea of the Finite Cell Method.}
		\label{fig:FCMIdea}
	\end{figure}
	
 First, an arbitrary complex shape defined on a physical domain $\Omega$ is immersed in a simplified box-like domain $\Omega_{e}$. Due to its simplicity, $\Omega_{e}$ can be trivially discretized with a structured grid of cuboids, further referred to as \textit{finite cells}. These elements provide the support for shape functions which are chosen to be integrated Legendre polynomials of order $p$.
	
Second, the original boundary value problem must be recovered on the actual, physical domain. To achieve such a result, an indicator function $\alpha(\bm{x})$ is introduced into the problem formulation. It is defined to be equal to one on all points of the physical domain $\Omega$ and to a small positive value in the domain $\Omega_e\backslash\Omega$. Then, the modified linear elastic weak form of the problem can be written as follows:
	
	\begin{equation}
	\begin{split}
	\text{Find } u_i(x_j)\in H_{\hat{u}}^1(\Omega_e) \text{ satisfying}&\\
	\int_{\Omega_e} \alpha(x_i)C_{ijkl}\frac{\partial u_k}{\partial x_l} \frac{\partial \delta v_i}{\partial x_j} \, d\Omega_{e} &+ \beta_D \int_{\Gamma_D} u_i \delta v_i \,d\Gamma_D =\\
	&= \int_{\Omega_e} \alpha(x_i) b_i\delta v_i d\Omega_{e} + \int_{\Gamma_N} \hat{t}_i \delta v_i \,d\Gamma_N  + \beta_D \int_{\Gamma_D} \hat{u}_i \delta v_i \,d\Gamma_D 
	\end{split}
	\label{eq:weakFormulationFCM}
	\end{equation}

with $ H_{\hat{u}}^1(\Omega_e)$ being the first-order Sobolev space, $\hat{u}$ indicating a prescribed displacement on the domain boundary $\Gamma_D$, and $\hat{t}$ is prescribed traction on boundary $\Gamma_N$. In the present work, Dirichlet boundary conditions are enforced using the penalty method with the penalty parameter $\beta_D$.
	
As the geometries under consideration stem from CT images, the spatial scalar function $\alpha(\bm{x})$ can be conveniently related to the acquired Hounsfield scale. Since the analyzed parts are metallic lattices, the contrast between material and void in the scan is commonly very high.

Therefore, the threshold value of Hounsfield units $HU_{thres}$ used to identify the metal and void regions in the CT scan images can directly be used to define the indicator function as follows:
	\begin{equation}
		\alpha(\bm{x})=\begin{cases} 1 \qquad\qquad\,\,\,\quad\text{ if } HU\geq HU_{thres}\\ \varepsilon, \varepsilon<<1 \qquad \text{ if } HU< HU_{thres}
		\end{cases}
		\label{eq:HUIndicator}
	\end{equation}

Finally, as the indicator function makes the domain integrands in~\cref{eq:weakFormulationFCM} discontinuous over the boundaries of the physical domain, a special integration rule should be applied. For this purpose, multiple techniques have been proposed (see e.g.~\cite{Abedian2013, Kudela2016}). However, the most efficient integration rule for CT-based geometrical models is a voxel-based pre-integration introduced in~\cite{Yang2012a}. In this case, the shape of an object is fully described by a grid of voxels with a constant Hounsfield scale. Such an underlying structure allows to further decompose every finite cell into a number of voxels $m_x \times m_y \times m_z$. Then, the standard $(p+1)^3$ quadrature rule can be applied to every voxel resulting in $m_x(p+1) \times m_y(p+1) \times m_z(p+1)$ integration points for one finite cell. Using these integration points, the integrands in~\cref{eq:weakFormulationFCM} can be efficiently pre-computed for every voxel in an offline phase. Then, the resulting matrices are scaled in an online stage with the indicator function $\alpha(\bm{x})$ as in~\cref{eq:HUIndicator}. Thus, this integration method provides an accurate and efficient technique to accurately compute the discontinuous integrands for CT-based geometries.
	
Although the Finite Cell Method in combination with a voxel-based pre-integration technique provides a powerful tool to perform numerical analysis directly on CT images, the size of the computed systems remains large. Large linear systems occur because as-manufactured structures include a considerable number of small-scale features, which are significant for the overall behavior of the parts. As an example, the largest CT scan considered further in this paper has a resolution of $2096\times272\times128$ voxels, while the smallest significant geometrical variations have a size of $3-7$ voxels. To capture this behavior a relatively fine FCM mesh must be employed, thus, leading to a large number of degrees of freedom. An appropriate way to handle these large scale computational systems is to use a hybrid parallelization technique as the one introduced by Jomo \textit{et al.} in~\cite{Jomo2016, Jomo2019}.

}
\section{Classical and strain gradient beam theories for uniaxial bending}
\label{sec:strainGradientTheory}
{
	Although one can expect that 3D numerical simulations on the as-manufactured AM lattices provide an accurate and realistic solution of the complex mechanical behavior, often a fast prediction is important for an early analysis stage. One of the approaches to obtain a quick solution is to use beam theories.

	When slender beams with a small thickness-to-length ratio are considered, an Euler-Bernoulli model can be used to evaluate bending rigidity, while the Timoshenko beam theory is more appropriate when shear effects are not negligible. Both Euler-Bernoulli and Timoshenko beam models rely on the determination of the effective Young's modulus $E\strut^{\ast}$, the moment of inertia $I\strut^{\ast}$, and the effective shear modulus $G\strut^{\ast}$ for the latter model. These three quantities are not straightforward to obtain when lattice structures are considered. The two most common ways to determine them are to perform experiments or to use a first-order numerical homogenization. The former will be considered in this paper for the determination of the as-manufactured effective Young's modulus via a tensile test, while the latter is used for the determination of as-designed effective quantities and as-manufactured effective shear modulus $G\strut^{\ast}$. For a detailed description of the first-order CT-based homogenization employed in this article, interested readers are referred to~\cite{Korshunova2020}. However, when the size effects in the material characterization of lattice structures under bending play an important role in the macroscopic response, the classical beam theories might deliver incorrect results and must be further enhanced, e.g., by means of high-order models such as the strain gradient beam theory described in the following.
	
	\subsection{Three-point bending problem of lattice beams}
	In the present work, three-point bending of the AM lattice beams is investigated. The structure deforms in the \textit{xz}-plane (see the 2D sketch of the problem in~\cref{fig:Bendign2DSketch}).
	
	\begin{figure}[H]
		\centering
		\hspace*{-.8cm}
		\def\svgwidth{\textwidth}
		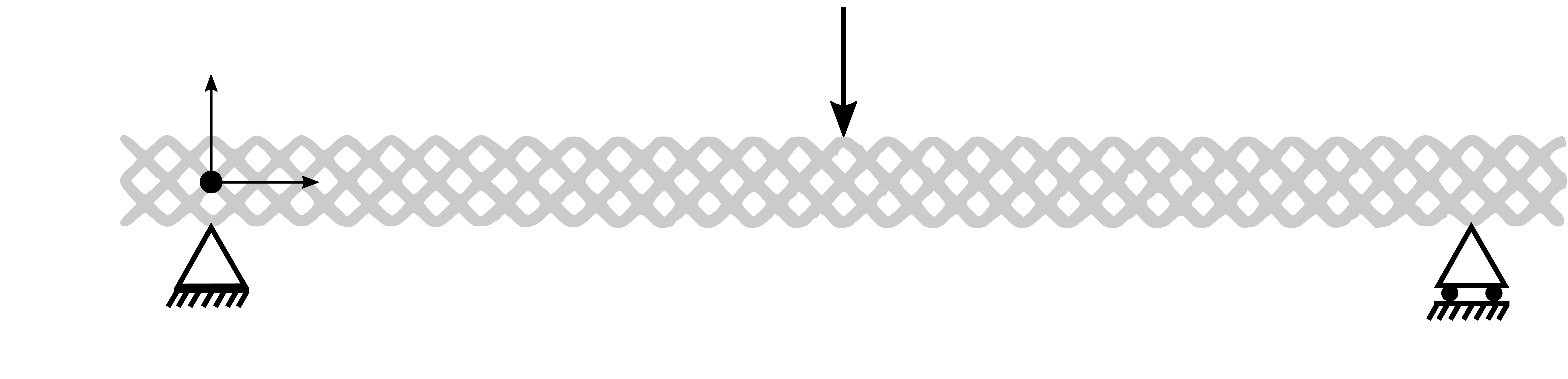
		\caption{A 2D sketch of a three-point bending setup.}
		\label{fig:Bendign2DSketch}
	\end{figure}

	The boundary conditions for this test:
	\begin{equation}
		w(x=0) = 0, \qquad
		M(x=0) = 0,\qquad
		w'\left(x=\frac{L}{2}\right)= 0, \qquad
		Q\left(x=\frac{L}{2}\right) = \frac{F}{2} 	 
		\label{eq:beamBCs}
	\end{equation}	
	where the $x-$coordinate runs along the central (neutral) axis of the beam and $x=0$ is the coordinate of a fixed support, $w$ is the deflection of a central axis of the beam, $F$ is the applied force at the middle of the beam span with respect to which the beam problem is symmetric, $M$ is the standard bending moment and $Q$ is the shear force in the beam.
	
	\subsection{Classical beam theories}	
	Given the previously defined bending problem, the classical Euler-Bernoulli solution delivers the maximum deflection at $x = \frac{L}{2}$:
	\begin{equation}
	w^{EB}=\frac{FL^3}{48E\strut^{\ast}I}
	\label{eq:solutiontEB}
	\end{equation}	
	where $L$ is the length of the beam, $E\strut^{\ast}$ is the effective Young's modulus. In the present work, we perform the homogenization such that the beam becomes a solid block made of homogeneous material with $E\strut^{\ast}$. Hence, $I$ is defined as a standard moment of inertia of a cross-section having the outer dimensions of the original structure.
	
	To account for shear deformations for higher thickness-to-length ratios, the solution of classical Timoshenko beam theory for three-point bending can be formulated as follows:
	\begin{equation}
	w^{T}=\frac{FL^3}{4E\strut^{\ast}I} + \frac{FL}{4G\strut^{\ast}A}
	\label{eq:solutiontT}
	\end{equation}
	\noindent where $G\strut^{\ast}$ is the effective shear modulus and $A$ is the cross-sectional effective area. Then, the main characteristic of the bending behavior is the bending stiffness or bending rigidity. It defines the resistance of the specimens to bending deformations and is determined as follows:
	\begin{equation}
	D = \frac{F}{w}
	\label{eq:bendingRigidity}
	\end{equation}	
	\noindent where $F$ is the applied load and $w$ the determined displacement.
	
	With the help of the classical beam theories solutions, this quantity can be determined analytically when all other parameters are known. The classical Euler-Bernoulli bending rigidity for the considered problem can be written as follows:
	
	\begin{equation}
	D^{EB}=\frac{F}{w^{EB}}=\frac{48E\strut^{\ast}I}{L^3}=\frac{4E\strut^{\ast}bh^3}{L^3}
	\label{eq:solutionrigidityEB}
	\end{equation}	
	
	where $b$ is the depth and $h$ is the thickness of the homogenized rectangular cross-section.
	
	Analogously, the classical bending rigidity using the Timoshenko beam theory is defined as:
	
	\begin{equation}
	D^{T}=\frac{F}{w^{T}}=\frac{D^{EB}}{1+\dfrac{12E\strut^{\ast} I}{G\strut^{\ast}AL^2}}=\frac{D^{EB}}{1+\dfrac{E\strut^{\ast}}{G\strut^{\ast}}\left(\dfrac{h}{L}\right)^2}
	\label{eq:solutionrigidityT}
	\end{equation}	
	
	\Cref{eq:solutionrigidityT} shows that for a fixed length $L$ the bending rigidity $D^T$ approaches $D^{EB}$ when thickness approaches zero, whereas for constant thickness-to-length ratios the Timoshenko and Euler-Bernoulli rigidities stay apart.

	\subsection{Strain gradient beam theory}
	 In the scope of the present work, we also consider strain gradient beam theories elaborated in~\cite{Khakalo2018, Niiranen2019}. 
	
	In the following, the derivation for the Euler-Bernoulli beam is described in greater detail.
	The strain energy density for a 3D body following Mindlin's strain gradient elasticity theory of form II is formulated as follows~\cite{Mindlin1968}:
	\begin{equation}
	\mathcal{W_{II}} = \frac{1}{2}C_{ijkl}\varepsilon_{ij}\varepsilon_{kl} +\frac{1}{2}A_{mijnkl}\partial_{m}\varepsilon_{ij}\partial_{n}\varepsilon_{kl}
	\end{equation}	
	where $C_{ijkl}$ and $A_{mijnkl}$ stand for the linear and high-order elasticity tensors, $\varepsilon_{ij}$ is the engineering strain tensor, and $\partial_{m}\varepsilon_{ij}$ and $\partial_{n}\varepsilon_{kl}$ denote the partial strain gradient. 
	Following the assumption of weak non-locality for isotropic materials \cite{Lazar2015}, the high-order elasticity tensor can be further simplified:
	\begin{equation}
	A_{mijnkl} = g^2\delta_{mn} C_{ijkl}
	\end{equation}
	where $g$ is an intrinsic length scale parameter affecting the macroscopic behavior and $\delta_{mn}$ is the Kroenecker delta. Parameter $g$ can be interpreted as a high-order material parameter for a specific microstructure. 
	 
	Using the principal of virtual work, the variation of the internal energy takes the form 
	\begin{equation}
	\delta\int_{\Omega} \mathcal{W_{II}}d\Omega = \int_{\Omega} \left( C_{ijkl}\varepsilon_{kl}\delta\varepsilon_{ij} + g^2\delta_{mn}C_{ijkl} \partial_{l}\varepsilon_{kl}\partial_k\delta\varepsilon_{ij} \right) d\Omega
	\label{eq:internalEnergyVirtual}
	\end{equation}
	where $\delta$ indicated the variation.
	
	Then, the dimensional reduction to the strain gradient Euler-Bernoulli beam theory is performed. The displacement components $\bm{u} = \left(u_x,u_y, u_z\right)$ obey the same relationships as for the classical beam theory:
	\begin{equation}
	u_x=-z\frac{\partial w(x)}{\partial x}, \qquad u_y=0, \qquad u_z = w(x)
	\label{eq:displacementComponents}
	\end{equation}
	where $x$ is the coordinate along the main axis of the beam, $z$ is the direction perpendicular to it, and $y$ is the out-of-plane coordinate, as depicted in \cref{fig:Bendign2DSketch}. This leaves the transverse deflection $w$ as the only unknown. 
	
	Furthermore, the only non-zeros stress and strain components are $\sigma_{xx}$ and $\varepsilon_{xx}$.
	With this background, the formulation of a generalized moment $R(x)$ can be introduced:
	\begin{equation}
	R(x) = \int_A \frac{\partial \sigma_{xx}(x,y,z)}{\partial z} dA
	\label{eq:moments}
	\end{equation}
	where $A = A(x)$ is the cross-sectional area of the beam.
	
	Then, the variation of the internal energy in~\cref{eq:internalEnergyVirtual} with~\cref{eq:displacementComponents} and~\cref{eq:moments} simplifies to the 1D energy expression over the main axis of the beam:
	\begin{equation}
	\delta\int_{\Omega} \mathcal{W_{II}}d\Omega = \int_{0}^{L} \left( M + g^2 R\right)\frac{\partial^2 (\delta w)}{\partial x^2} dx + \int_{0}^{L} g^2 \frac{\partial M}{\partial x} \frac{\partial^3(\delta w)}{\partial x^3}
	\end{equation} 
	Applying the Hamilton's principle the strong formulation of the one-parameter strain gradient Euler-Bernoulli elasticity model can then be formulated:
	\begin{equation}
	(M + g^2R-(g^2 M')')''=f \qquad \forall x\in(0,L)
	\end{equation}	
	where $f$ is the externally applied force, $g$ is an unknown high-order material parameter, and the high-order term $((g^2 M')')''$ is responsible for the description of boundary layer effects. As the macroscopic behavior of the beam is of interest, the strong form of the governing equation for constant homogenized parameters can be further simplified: 
	\begin{equation}
 	( E\strut^{\ast}I + E\strut^{\ast}Ag^2)w''''=f \qquad  \forall x\in(0,L)
 	\label{eq:strongFormstrainGradientEB}
	\end{equation}

	The analytical solution of ~\cref{eq:strongFormstrainGradientEB} under the absence of body load with the boundary conditions described in~\cref{eq:beamBCs} takes the form:
	\begin{equation}
	w_{gr}^{EB}=\frac{FL^3}{48 \left(E\strut^{\ast}I+E\strut^{\ast}Ag^2\right)}
	\label{eq:solutionGradientEB}
	\end{equation}

	\Cref{eq:solutionGradientEB} compared to the solution of the classical Euler-Bernoulli theory in~\cref{eq:solutiontEB} introduces the intrinsic length scale parameter $g$ which acts as a high-order material parameter depending on the microstructure of the unit cell. This parameter characterizes the size-dependent beam behavior when the thinnest beams show a stiffening effect.
	
	The solution of the strain gradient Timoshenko beam theory can be derived in a similar manner taking into account the respective assumptions:
	
	\begin{equation}
	w_{gr}^T = \frac{FL^3}{48 \left(E\strut^{\ast}I+E\strut^{\ast}Ag^2\right)} + \frac{FL}{4G\strut^{\ast}A}
	\label{eq:solutionGradientT}
	\end{equation}
	
	\Cref{eq:solutionGradientT} is also similar to the solution of the classical Timoshenko theory except for the presence of the intrinsic material parameter $g$. The bending rigidities (with rectangular cross sections $A = bh$) corresponding to these deflections can be shown to follow, respectively, the formulae:
	\begin{equation}
	\begin{aligned}
	D_{gr}^{EB} &= D^{EB}\left(1 + 12\left(\frac{g}{h}\right)^2\right)\\
	D_{gr}^{T} &= D^T\left(1 + 12\left(\frac{g}{h}\right)^2\right)
	\label{eq:strainGradientRigidity}
	\end{aligned}
	\end{equation}
	revealing the size effect for decreasing values of $h$ with a fixed value of $g$.
	
	To sum up, both the classical and the strain-gradient theories could provide a quick estimate of the bending behavior of the considered beam-like lattice structures. In the following, the predictions provided by these theories will be compared to the full 3D numerical and experimental analysis performed on the AM octet-truss beams. Furthermore, their accuracy and applicability will be evaluated with the help of experimental three-point bending tests.
}

\newcommand{\pictureDir}{./sections/numerics/Pictures}

%% file: sections/numerics/Pictures/PhysicalDomain.pdf_tex
\begingroup%
  \makeatletter%
  \providecommand\color[2][]{%
    \errmessage{(Inkscape) Color is used for the text in Inkscape, but the package 'color.sty' is not loaded}%
    \renewcommand\color[2][]{}%
  }%
  \providecommand\transparent[1]{%
    \errmessage{(Inkscape) Transparency is used (non-zero) for the text in Inkscape, but the package 'transparent.sty' is not loaded}%
    \renewcommand\transparent[1]{}%
  }%
  \providecommand\rotatebox[2]{#2}%
  \newcommand*\fsize{\dimexpr\f@size pt\relax}%
  \newcommand*\lineheight[1]{\fontsize{\fsize}{#1\fsize}\selectfont}%
  \ifx\svgwidth\undefined%
    \setlength{\unitlength}{1616.12952525bp}%
    \ifx\svgscale\undefined%
      \relax%
    \else%
      \setlength{\unitlength}{\unitlength * \real{\svgscale}}%
    \fi%
  \else%
    \setlength{\unitlength}{\svgwidth}%
  \fi%
  \global\let\svgwidth\undefined%
  \global\let\svgscale\undefined%
  \makeatother%
  \begin{picture}(1,0.57305404)%
    \lineheight{1}%
    \setlength\tabcolsep{0pt}%
    \put(0,0){\includegraphics[width=\unitlength,page=1]{PhysicalDomain.pdf}}%
    \put(0.472,0.265){\color[rgb]{0,0,0}\makebox(0,0)[lt]{\lineheight{1.25}\smash{\begin{tabular}[t]{l}$\Omega$\end{tabular}}}}%
  \end{picture}%
\endgroup%

%% file: sections/numerics/Pictures/ExtendedDomain.pdf_tex
\begingroup%
  \makeatletter%
  \providecommand\color[2][]{%
    \errmessage{(Inkscape) Color is used for the text in Inkscape, but the package 'color.sty' is not loaded}%
    \renewcommand\color[2][]{}%
  }%
  \providecommand\transparent[1]{%
    \errmessage{(Inkscape) Transparency is used (non-zero) for the text in Inkscape, but the package 'transparent.sty' is not loaded}%
    \renewcommand\transparent[1]{}%
  }%
  \providecommand\rotatebox[2]{#2}%
  \newcommand*\fsize{\dimexpr\f@size pt\relax}%
  \newcommand*\lineheight[1]{\fontsize{\fsize}{#1\fsize}\selectfont}%
  \ifx\svgwidth\undefined%
    \setlength{\unitlength}{1616.12952525bp}%
    \ifx\svgscale\undefined%
      \relax%
    \else%
      \setlength{\unitlength}{\unitlength * \real{\svgscale}}%
    \fi%
  \else%
    \setlength{\unitlength}{\svgwidth}%
  \fi%
  \global\let\svgwidth\undefined%
  \global\let\svgscale\undefined%
  \makeatother%
  \begin{picture}(1,0.57305404)%
    \lineheight{1}%
    \setlength\tabcolsep{0pt}%
    \put(0,0){\includegraphics[width=\unitlength,page=1]{ExtendedDomain.pdf}}%
    \put(0.463,0.262){\color[rgb]{0,0,0}\makebox(0,0)[lt]{\lineheight{1.25}\smash{\begin{tabular}[t]{l}$\Omega$\end{tabular}}}}%
    \put(0.39689347,0.50603044){\color[rgb]{0,0,0}\makebox(0,0)[lt]{\lineheight{1.25}\smash{\begin{tabular}[t]{l}$\Omega_e\backslash\Omega$\end{tabular}}}}%
  \end{picture}%
\endgroup%

%% file: sections/numerics/Pictures/InidcatorMesh.pdf_tex
\begingroup%
  \makeatletter%
  \providecommand\color[2][]{%
    \errmessage{(Inkscape) Color is used for the text in Inkscape, but the package 'color.sty' is not loaded}%
    \renewcommand\color[2][]{}%
  }%
  \providecommand\transparent[1]{%
    \errmessage{(Inkscape) Transparency is used (non-zero) for the text in Inkscape, but the package 'transparent.sty' is not loaded}%
    \renewcommand\transparent[1]{}%
  }%
  \providecommand\rotatebox[2]{#2}%
  \newcommand*\fsize{\dimexpr\f@size pt\relax}%
  \newcommand*\lineheight[1]{\fontsize{\fsize}{#1\fsize}\selectfont}%
  \ifx\svgwidth\undefined%
    \setlength{\unitlength}{1597.28217658bp}%
    \ifx\svgscale\undefined%
      \relax%
    \else%
      \setlength{\unitlength}{\unitlength * \real{\svgscale}}%
    \fi%
  \else%
    \setlength{\unitlength}{\svgwidth}%
  \fi%
  \global\let\svgwidth\undefined%
  \global\let\svgscale\undefined%
  \makeatother%
  \begin{picture}(1,0.58473184)%
    \lineheight{1}%
    \setlength\tabcolsep{0pt}%
    \put(0,0){\includegraphics[width=\unitlength,page=1]{InidcatorMesh.pdf}}%
    \put(0,0){\includegraphics[width=\unitlength,page=2]{InidcatorMesh.pdf}}%
    \put(0.19987323,0.25002339){\color[rgb]{0,0,0}\makebox(0,0)[lt]{\lineheight{1.25}\smash{\begin{tabular}[t]{l}$\scriptstyle \bm{\alpha\approx0}$\end{tabular}}}}%
    \put(0.42513438,0.2987379){\color[rgb]{0,0,0}\makebox(0,0)[lt]{\lineheight{1.25}\smash{\begin{tabular}[t]{l}$\scriptstyle \bm{\alpha=1}$\end{tabular}}}}%
  \end{picture}%
\endgroup%

%% file: sections/numerics/Pictures/Bending.pdf_tex
\begingroup%
  \makeatletter%
  \providecommand\color[2][]{%
    \errmessage{(Inkscape) Color is used for the text in Inkscape, but the package 'color.sty' is not loaded}%
    \renewcommand\color[2][]{}%
  }%
  \providecommand\transparent[1]{%
    \errmessage{(Inkscape) Transparency is used (non-zero) for the text in Inkscape, but the package 'transparent.sty' is not loaded}%
    \renewcommand\transparent[1]{}%
  }%
  \providecommand\rotatebox[2]{#2}%
  \newcommand*\fsize{\dimexpr\f@size pt\relax}%
  \newcommand*\lineheight[1]{\fontsize{\fsize}{#1\fsize}\selectfont}%
  \ifx\svgwidth\undefined%
    \setlength{\unitlength}{1712.73533618bp}%
    \ifx\svgscale\undefined%
      \relax%
    \else%
      \setlength{\unitlength}{\unitlength * \real{\svgscale}}%
    \fi%
  \else%
    \setlength{\unitlength}{\svgwidth}%
  \fi%
  \global\let\svgwidth\undefined%
  \global\let\svgscale\undefined%
  \makeatother%
  \begin{picture}(1,0.2415361)%
    \lineheight{1}%
    \setlength\tabcolsep{0pt}%
    \put(0,0){\includegraphics[width=\unitlength,page=1]{Bending.pdf}}%
    \put(0.19839203,0.11035896){\color[rgb]{0,0,0}\makebox(0,0)[lt]{\lineheight{1.25}\smash{\begin{tabular}[t]{l}$x$\end{tabular}}}}%
    \put(0.11700111,0.18998085){\color[rgb]{0,0,0}\makebox(0,0)[lt]{\lineheight{1.25}\smash{\begin{tabular}[t]{l}$z$\end{tabular}}}}%
    \put(0.54757756,0.22876699){\color[rgb]{0,0,0}\makebox(0,0)[lt]{\lineheight{1.25}\smash{\begin{tabular}[t]{l}$F$\end{tabular}}}}%
    \put(0,0){\includegraphics[width=\unitlength,page=2]{Bending.pdf}}%
    \put(0.52997744,0.02281892){\color[rgb]{0,0,0}\makebox(0,0)[lt]{\lineheight{1.25}\smash{\begin{tabular}[t]{l}$L$\end{tabular}}}}%
  \end{picture}%
\endgroup%

%% file: sections/experiments/experiments.tex
\section{Experimental setup}
\label{sec:experimentalTest}
{
	The experimental and numerical investigations are held on octet-truss lattices. A representative unit cell of such structures is depicted in~\cref{fig:unitCellOctet}. As the main focus of the present work is the investigation of lattice bending behavior, an octet-truss unit cell indicated in~\cref{fig:unitCellOctet} is used to construct the four beam-like structures shown in~\cref{fig:BeamModels}. These beams have the same length of 128 mm ($32$ cells) and the same width of 8 mm ($2$ cells) but different heights (thicknesses): 4, 8, 12, and 16 mm, respectively ($1,2,3$, and $4$ unit cells). Thus, the constructed thickness-to-length ratios are $0.03$, $0.06$, $0.09$, and $0.13$ respectively.
	
	\begin{figure}[H]
		\centering
		\includegraphics[scale=0.26]{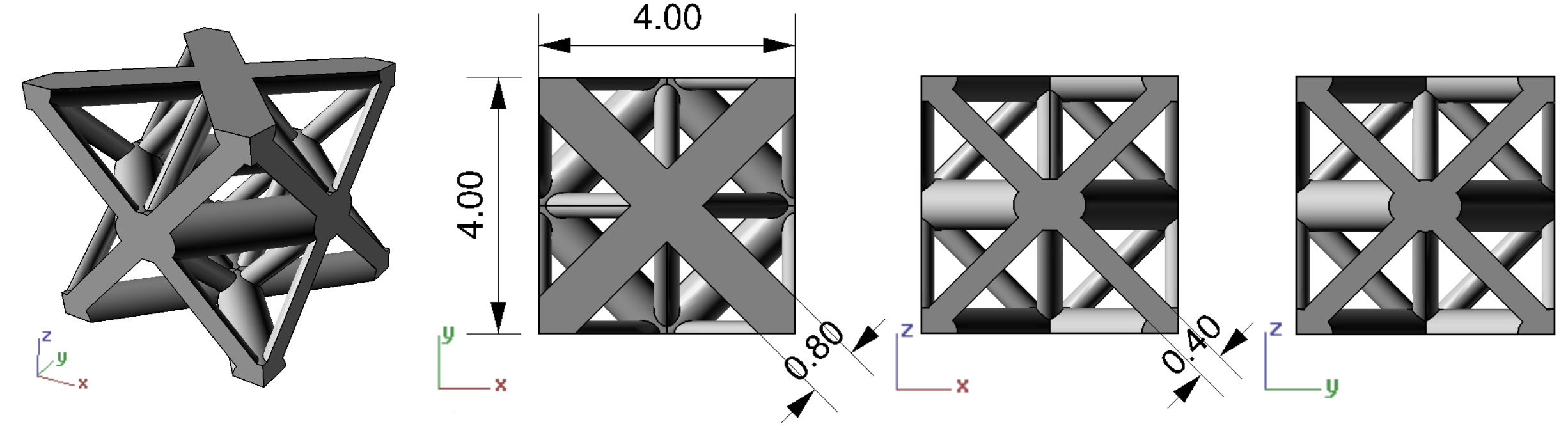}
		\caption{CAD model of the octet-truss unit cell~\cite{Korshunova2020a}.}
		\label{fig:unitCellOctet}
	\end{figure}

	\begin{figure}[H]
		\captionsetup[subfigure]{labelformat=empty}
		\centering
		\subfloat[(a) Beam $2\times32\times1$ ]{\includegraphics[scale=0.17]{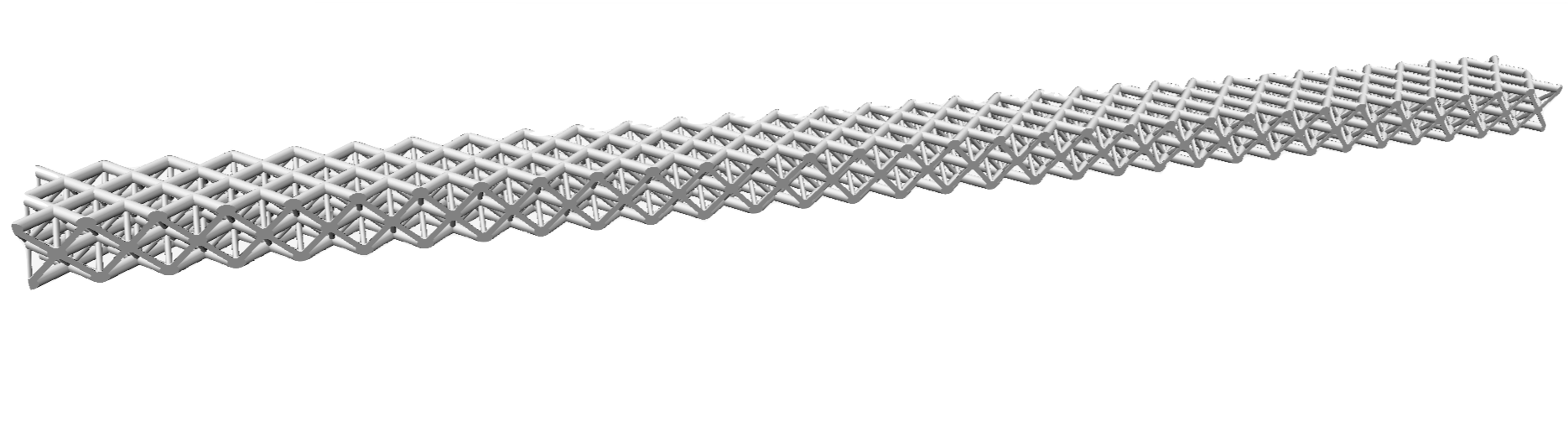}}
		\subfloat[(b) Beam $2\times32\times2$]{\includegraphics[scale=0.17]{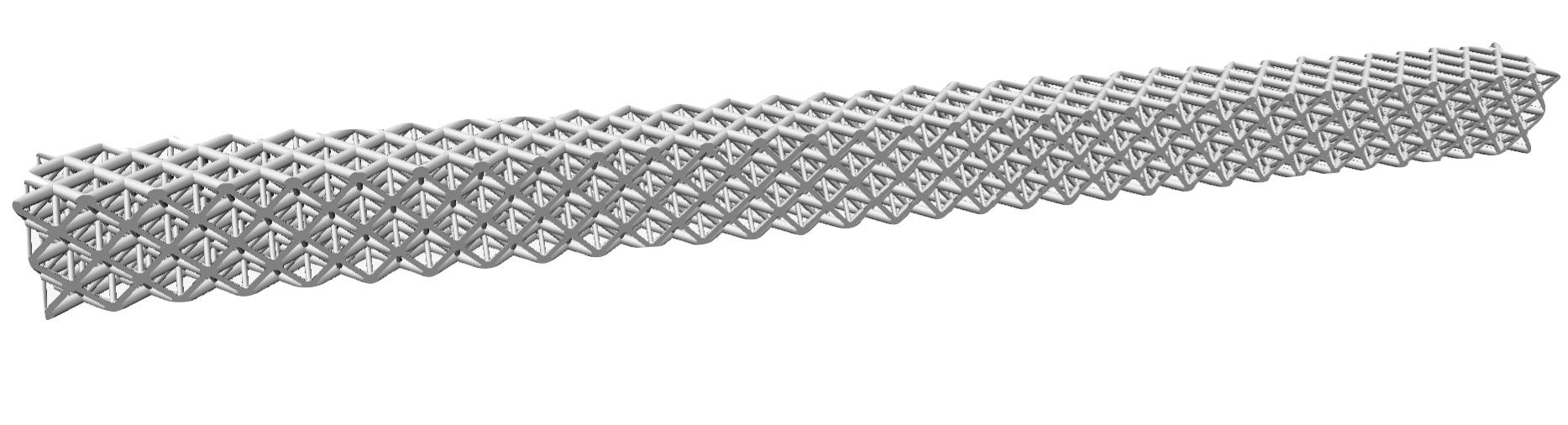}}\\
		\subfloat[(c) Beam $2\times32\times3$]{\includegraphics[scale=0.17]{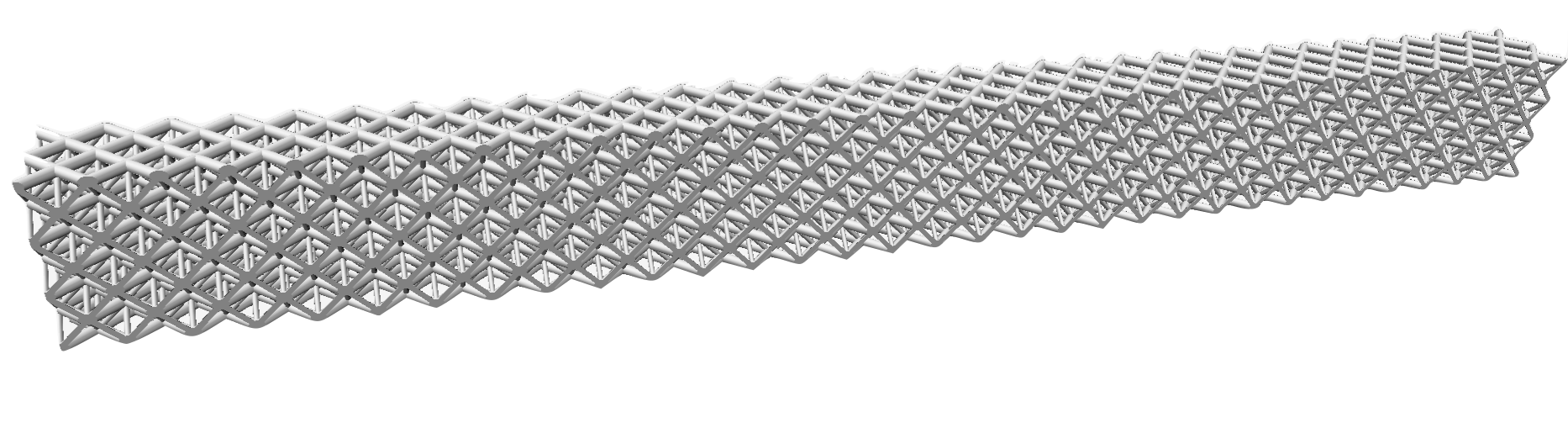}}
		\subfloat[(d) Beam $2\times32\times4$]{\includegraphics[scale=0.17]{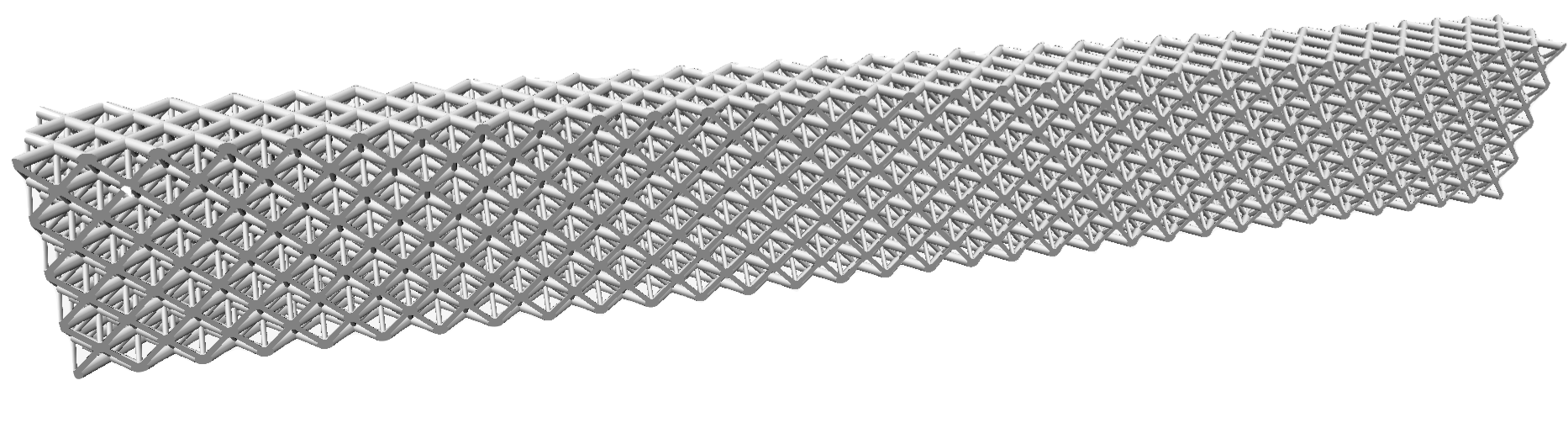}}
		\caption{Investigated CAD models of the octet-truss beam structures.}
		\label{fig:BeamModels}
	\end{figure}
	
	The specimens for experimental testing were printed in the laboratory 3DMetal@UniPV using a selective laser melting metal 3D printer Renishaw AM400. For the production of the specimens, stainless steel powder SS 316L-0407 was used. According to the material data sheet of the producer~\cite{ManualReinshaw}, the considered setup leads to a bulk material with Young's modulus $190\, \text{GPa} \pm 10 \text{GPa}$ in the printing direction. The produced specimens after heat treatment at $400^{\circ}$C in the chamber Nabertherm LH120/12 for 2 hours are shown in~\cref{fig:PrintedSpecimens}.

	\newcommand{\graphDir}{./sections/experiments/Pictures}
	
	\begin{figure}[H]
		\centering
		\includegraphics[scale=.1,trim={17cm 0cm 17cm 10cm},clip]{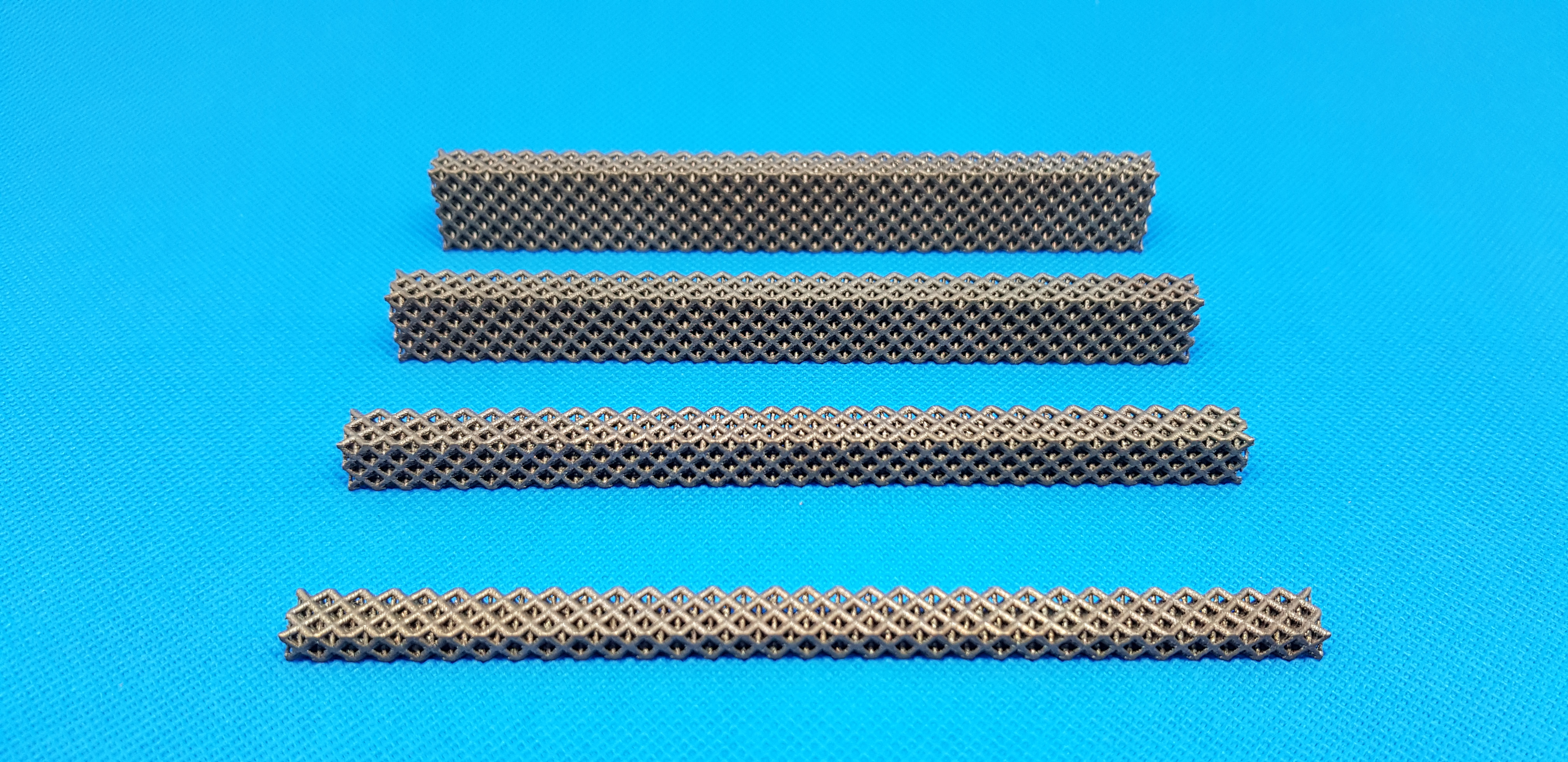}
		\caption{Printed specimens after heat treatment.}
		\label{fig:PrintedSpecimens}
	\end{figure}
	
	Prior to performing any experimental test, the four bending specimens were subjected to a computed tomography to acquire the as-manufactured geometries. The CT scans were performed with a Phoenix V CT scanner with a resolution of 61 $\mu $m. 
	
	Then, to validate the numerical frameworks proposed in~\cref{sec:NumericsFCM,sec:strainGradientTheory}, three main quantities were determined experimentally. These are the porosity of the printed lattices structures, the effective Young's modulus, and the bending rigidity.
	\paragraph{Porosity of the printed structure\newline}
	The overall porosity of the lattice structures is measured for two reasons. The first motivation is to compare the experimentally determined porosity value to the as-designed CAD-based ones, thus, providing the first estimate on the geometrical variations of the as-manufactured geometries with respect to the original CAD models shown in \cref{fig:BeamModels}. The second reason is to experimentally verify the porosity values determined from the acquired CT scan of every beam. The porosity values are determined by evaluating the mass of the specimen m. Then, considering the printed density $\rho$ indicated in~\cite{ManualReinshaw} the overall porosity can be calculated as:
	
	\begin{equation}
	\phi = 1 - \frac{m}{\rho V}
	\label{eq:experimentalPorosity}
	\end{equation} 
	
	where $V$ is the measured volume of the bounding box of the specimen.
	Together with the measured porosity values, the measurement uncertainty is computed based on the accuracy of the used instrumentation. 
	\paragraph{Effective Young's modulus\newline}	
	The second quantity of interest is the effective Young's modulus of the octet-truss lattice. This value is important for the investigation of the applicability of the beam models as described in~\cref{sec:strainGradientTheory}. The as-manufactured effective Young's modulus $E\strut^{\ast}$ is determined via a tensile test of the sample lattice specimens. The experiment is performed in the material mechanics laboratory with the help of the MTS Insight System. For the elongation measurements, a video extensometer is used (see experimental setup in~\cref{fig:TensileTest}). The effective Young's Modulus is then computed according to ASTM E111 standard~\cite{ASTME111}. The determined value is $E\strut^{\ast} = 12\,533\pm751$ MPa together with the corresponding measurement error.
	
	\begin{figure}[H]
		\centering
		\includegraphics[scale=0.07, trim={5cm 40cm 17cm 20cm},clip]{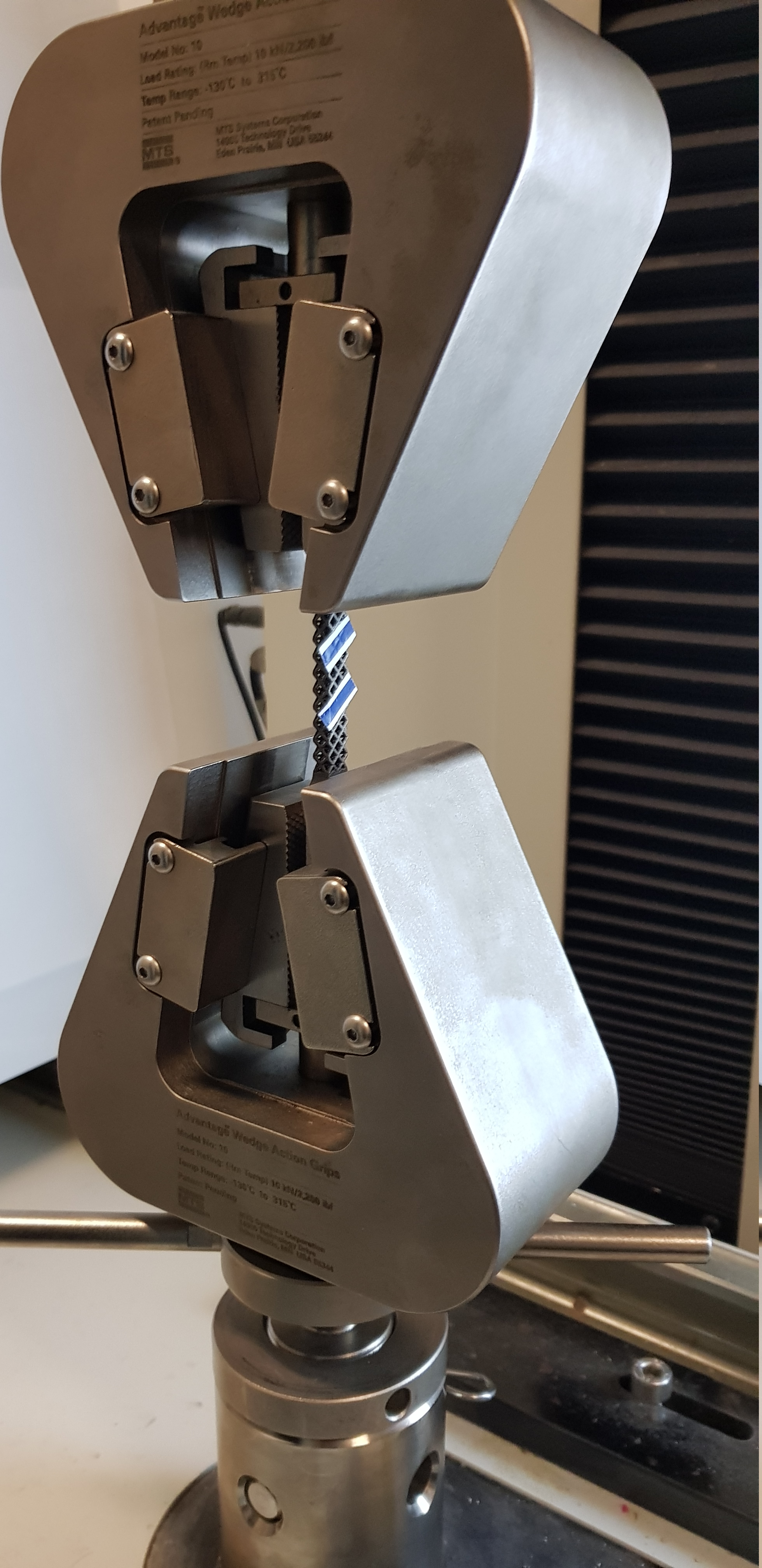}
		\caption{Experimental setup of a tensile experiment on an octet-truss lattice structure~\cite{Korshunova2020a}.}
		\label{fig:TensileTest}
	\end{figure}
	
	\paragraph{Bending rigidity\newline}	
	The final experimentally determined value is the bending rigidity of the octet-truss lattice beams as defined in~\cref{eq:bendingRigidity}. This quantity describes the characteristic overall (global) resistance of the structure against the bending deformation. The values of bending rigidity of the four 3D printed structures of \cref{fig:PrintedSpecimens} is experimentally measured by a three-point bending test under quasi-static conditions and displacement-controlled velocity (see~\cref{fig:BendingTest}). The span ($L$) between the supports is 120 mm, while the applied point load ($F$) is transferred in the middle of the span of the beam. During the experiment, the imposed displacement and the corresponding force are recorded. The bending rigidities of the beams are then computed by using~\cref{eq:bendingRigidity}. All tests are performed in both elastic and plastic regime. However, for the aim of this work only the elastic characteristics are considered. Experimental results will be discussed together with the numerical values in the following sections.
	
	\begin{figure}[H]
		\captionsetup[subfigure]{labelformat=empty}
		\centering
		\subfloat[(a) Beam $2\times32\times1$ ]{\includegraphics[scale=0.085,trim={30cm 0cm 40cm 10cm},clip]{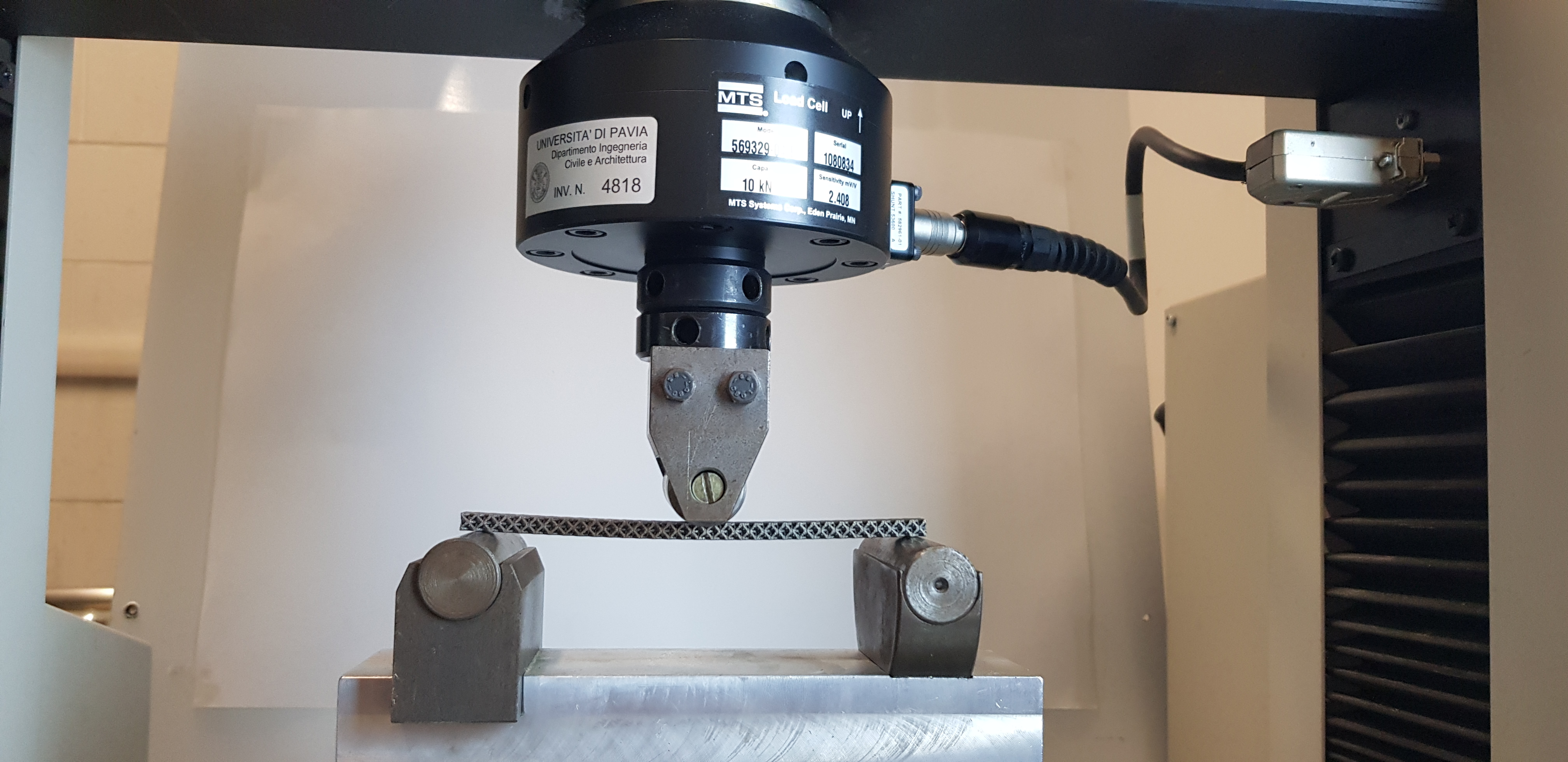}}
		\hspace*{1cm}
		\subfloat[(b) Beam $2\times32\times2$]{\includegraphics[scale=0.11,trim={46cm 18cm 40cm 5cm},clip]{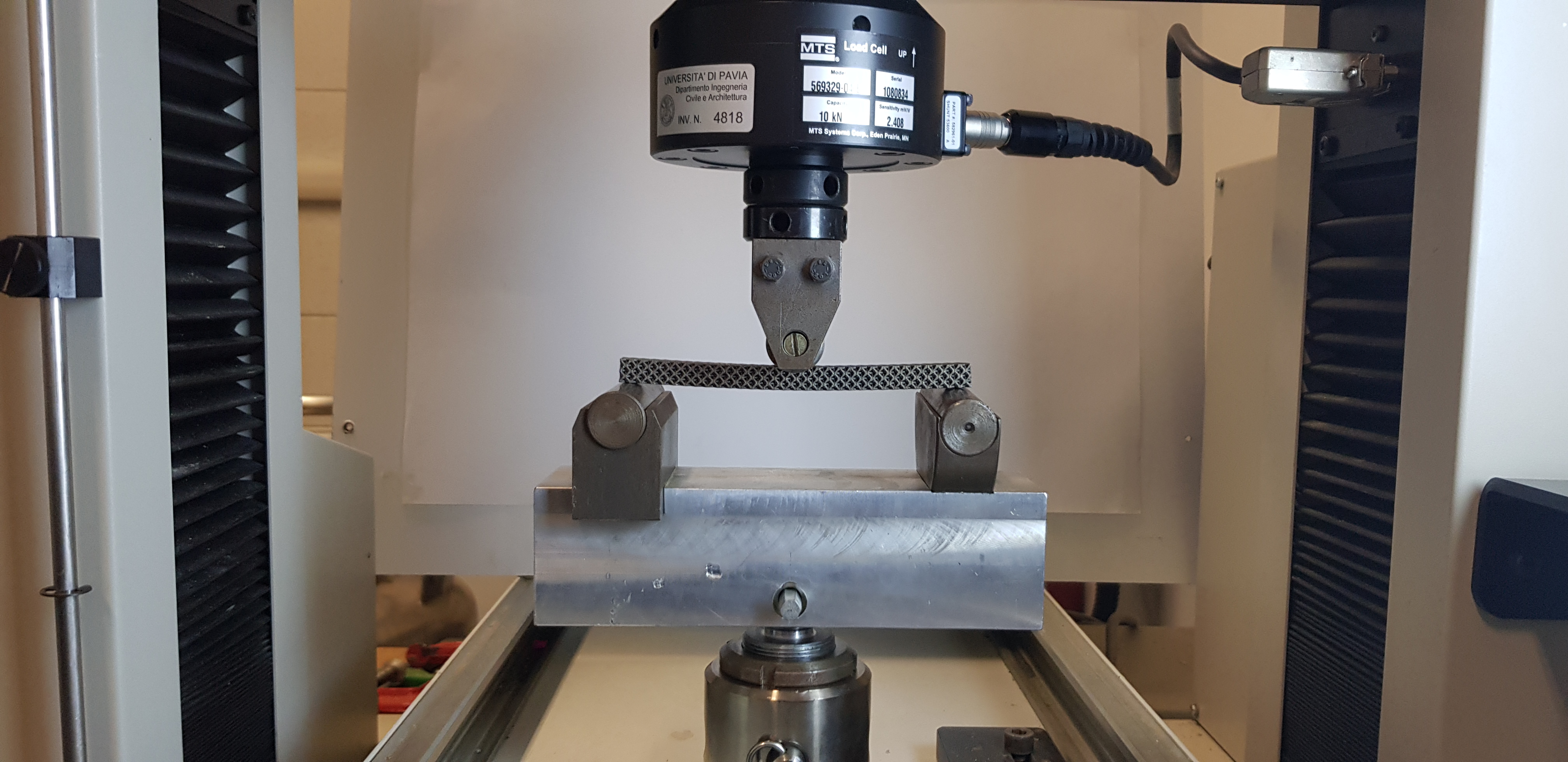}} \\
		\hspace*{0cm}\subfloat[(c) Beam $2\times32\times3$ ]{\includegraphics[scale=0.109,trim={45cm 13cm 40cm 13cm},clip]{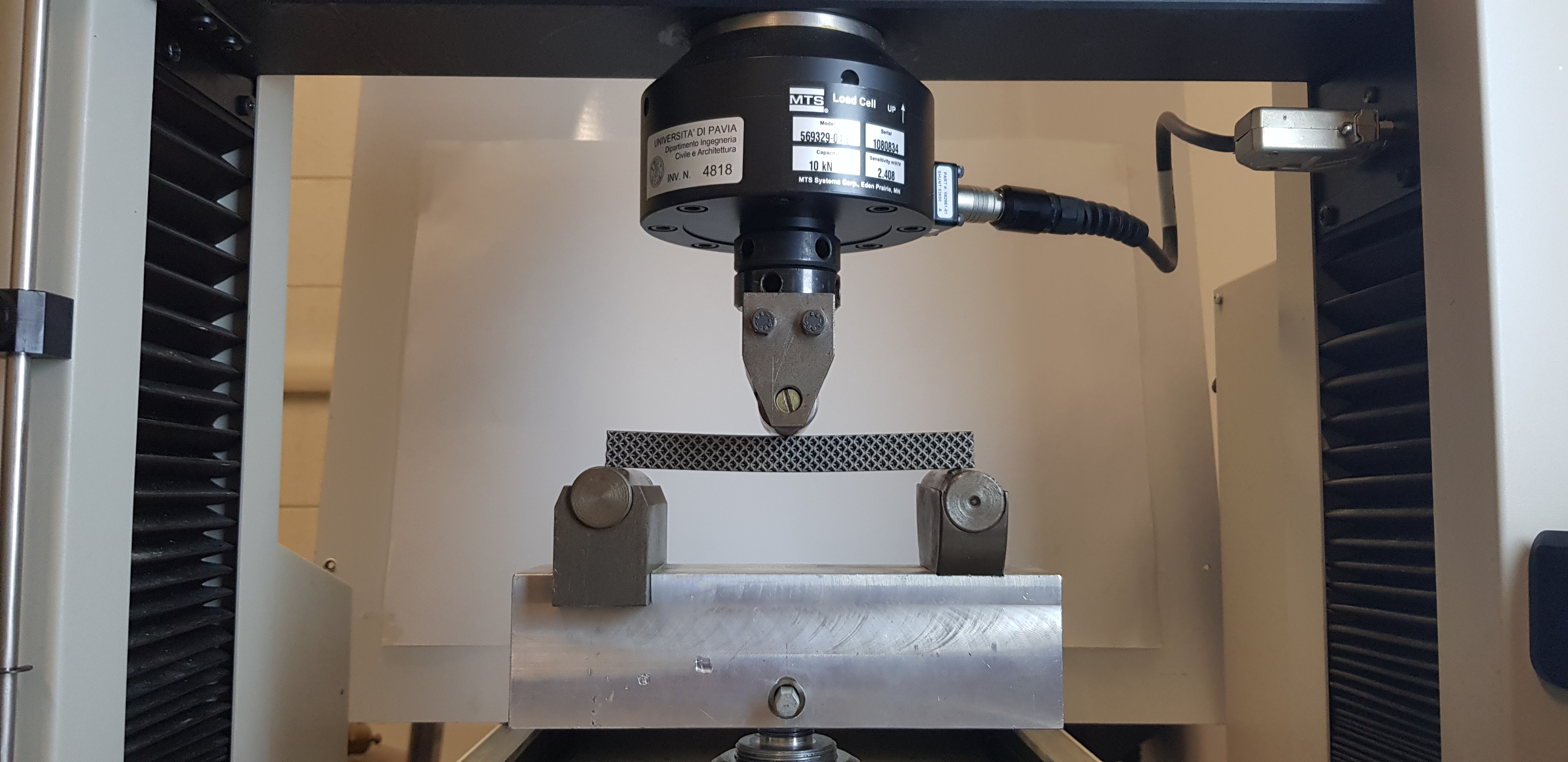}}
		\hspace*{1cm}
		\subfloat[(d) Beam $2\times32\times4$ ]{\includegraphics[scale=0.11,trim={45cm 26cm 40cm 0cm},clip]{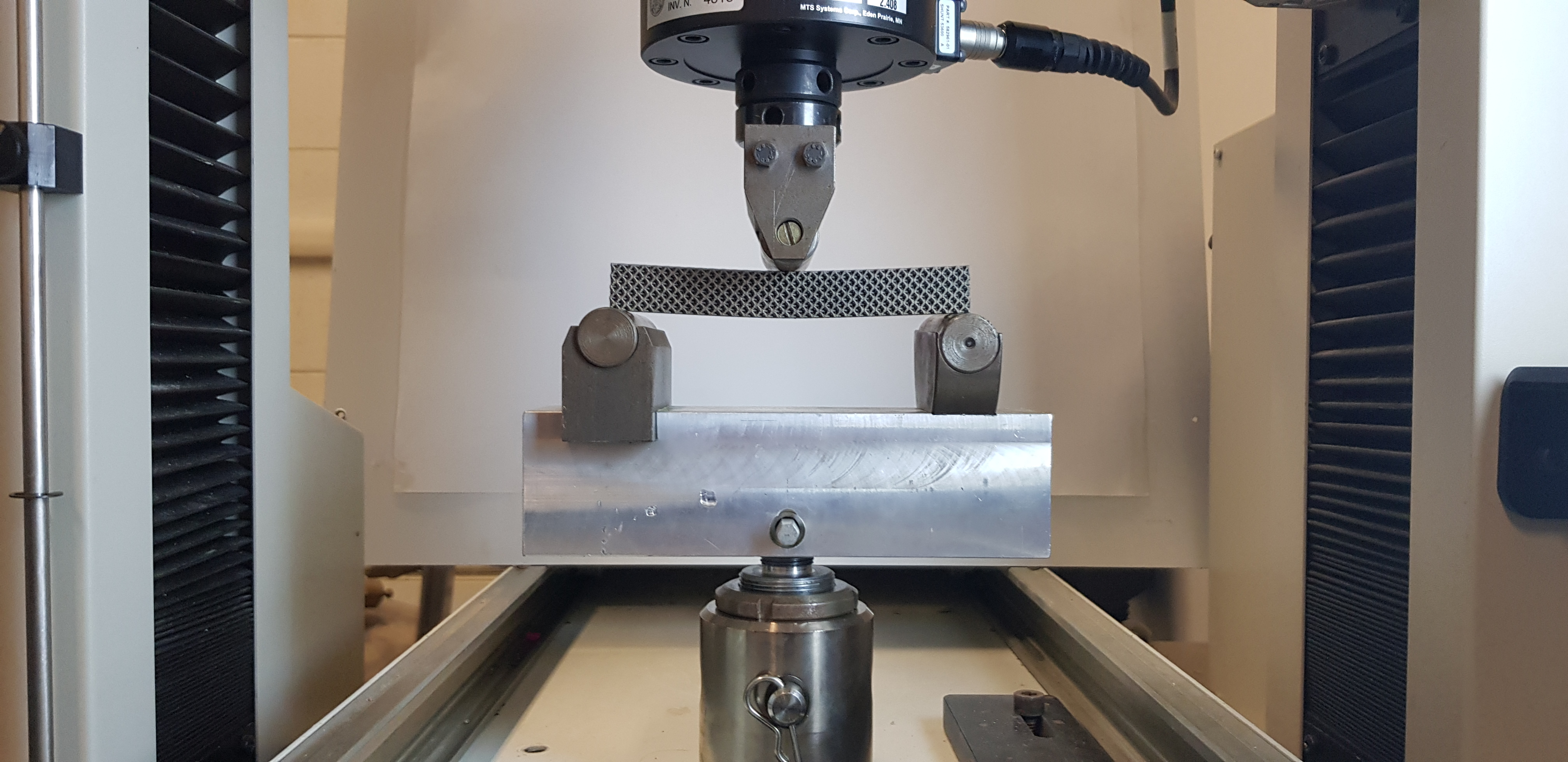}}
		\caption{Bending of beam specimens.}
		\label{fig:BendingTest}
	\end{figure}

}

%% file: sections/results/results.tex
\section{Numerical investigations}
\label{sec:numericalInvestigation}
{
	
	In this section, the results of the numerical investigations on the octet-truss lattices are discussed in detail.
	
	First, the behavior of the octet-truss lattice structures undergoing a bending load case is analyzed numerically in~\cref{subsec:bendingTest}. In this section, the as-manufactured and as-designed octet-truss beams are compared geometrically and the differences are quantified by means of the macroscopic porosity defined in~\cref{eq:experimentalPorosity}. Then, the direct numerical simulation of the three-point bending test is performed on both CAD and CT geometries. The achieved numerical results are finally compared to the experimental values.
	
	Second, in \cref{subsec:strainTheoryResults}, the applicability of the beam theories described in~\cref{sec:strainGradientTheory} is investigated. Both, the classical and the strain-gradient Euler-Bernoulli and Timoshenko beam theories are applied to analyze the behavior of both as-designed and as-manufactured octet-truss lattice beams.

	\subsection{Comparison of as-manufactured and as-designed mechanical behavior in bending}
	\label{subsec:bendingTest}
	{
		\newcommand{\graphDir}{./sections/experiments/Pictures}
		
		\paragraph{Geometrical comparison\newline}
		To highlight the macroscopic differences between the as-manufactured geometry extracted from CT scan images and the as-designed geometric model, zoomed views on both geometries are depicted in~\cref{fig:PrintedSpecimensComparisonCTCAD}. From a thorough comparison of the two geometric models (see \cref{fig:bendingSpecimenCombined}), the following geometrical features of as-manufactured geometry can be observed compared to the as-designed ones: 
		\begin{itemize}
			\item larger truss thickness;
			\item partially melted material powder particles in overhanging surfaces opposite to the build direction;
			\item excess material collection in the nodes.
		\end{itemize}
		These features are well-known side effects of the SLM printing process. It is also established in literature~\cite{Cao2020, Duplesis2020,Korshunova2020a}, that these geometrical features have a strong influence also on the numerical results, and thus as-designed models lead to a quite inaccurate prediction of the mechanical behavior of lattice structures. 
		
		\begin{figure}[H]
			\captionsetup[subfigure]{labelformat=empty}
			\centering
			\includegraphics[scale=0.4,trim={5cm 5cm 0cm 5cm},clip]{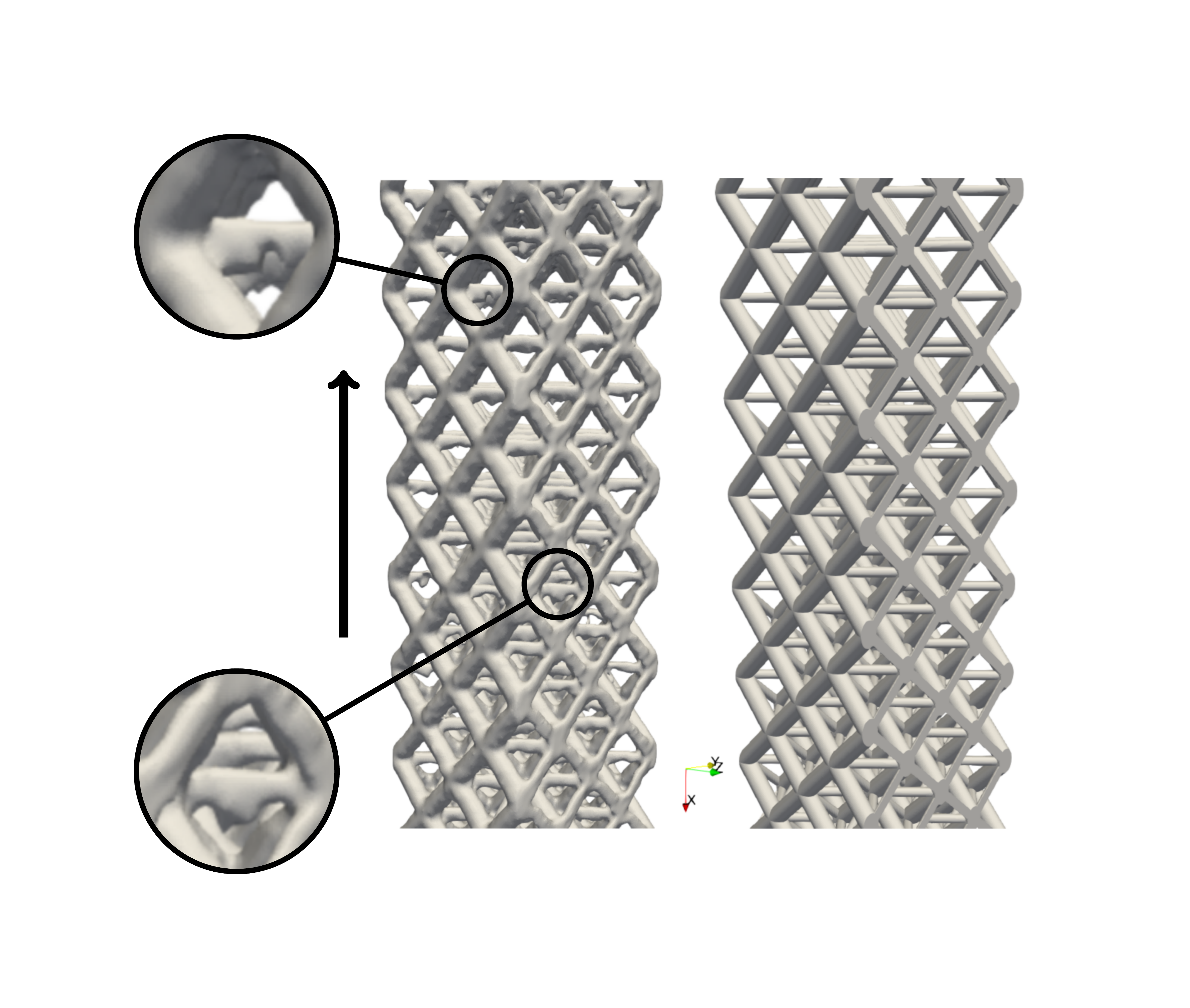}
			\caption{Zoom on the geometrical features of the as-manufactured (left) and as-designed (right) bending specimen (build direction marked with the black arrow).}
			\label{fig:PrintedSpecimensComparisonCTCAD}
		\end{figure}

		\begin{figure}[H]
			\captionsetup[subfigure]{labelformat=empty}
			\centering
			\includegraphics[scale=0.55,trim={4cm 5cm 4cm 5cm},clip]{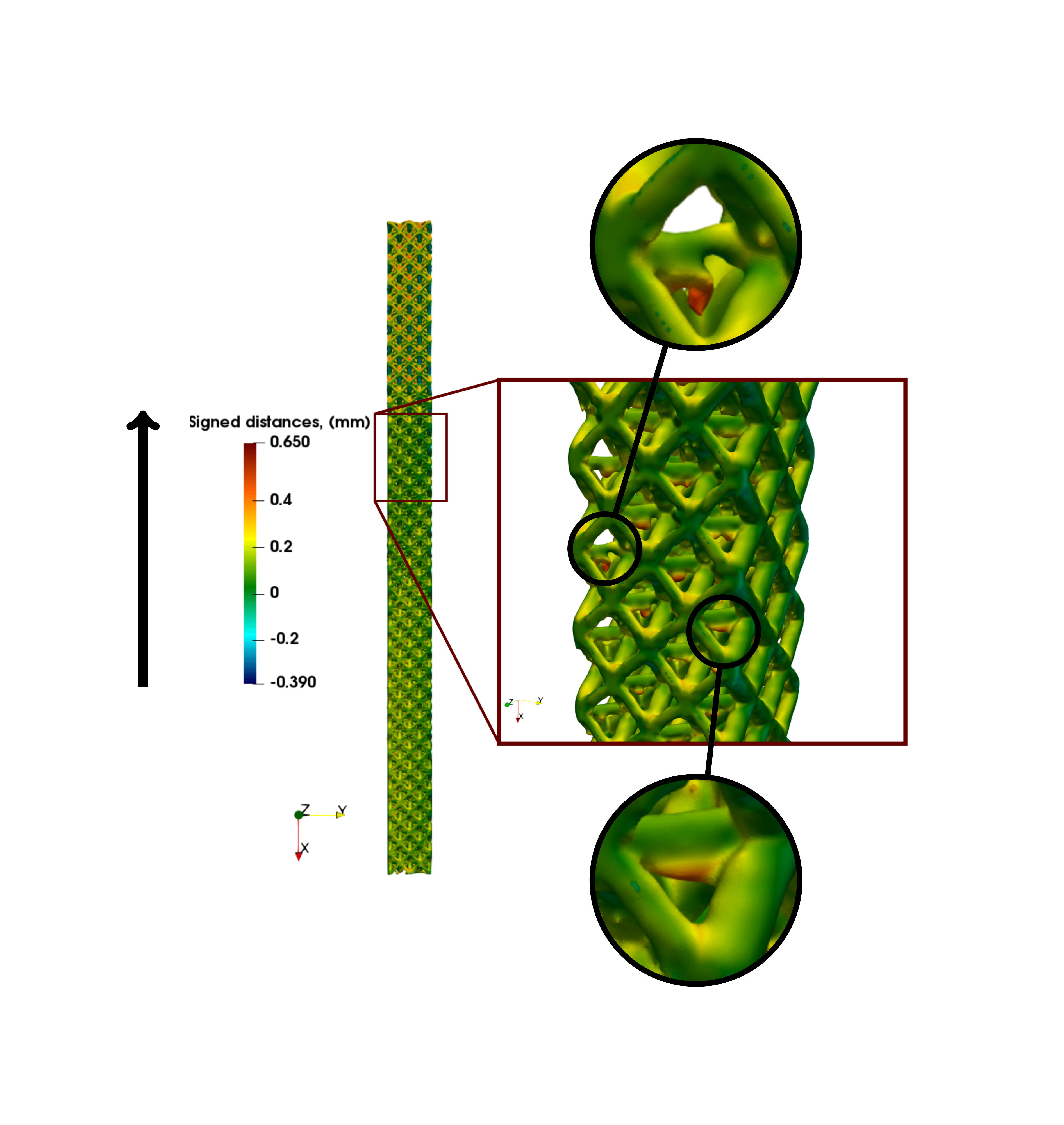}
			\caption{Comparison of as-manufactured and as-built octet-truss bending specimen $2\times32\times2$ (build direction marked with the black arrow).}
			\label{fig:bendingSpecimenCombined}
		\end{figure}
		
		Furthermore, as described in~\cref{sec:experimentalTest} the CT-based porosity values are compared to the CAD-based and experimental ones. \Cref{tab:PorosityBeamsTogetherWithCADAndCT} summarizes the achieved results. As expected, the CAD-based porosity is always larger than the printed one. This is also supported by the geometrical comparison of the CAD and CT-based model (an example of specimen $2\times32\times2$ is shown in~\cref{fig:bendingSpecimenCombined}). The excess material collection in the nodes together with the larger truss thickness leads to a lower manufactured porosity. Overall, the CT-based porosity is in good agreement with the experimental values, making us confident in the sufficient accuracy of the as-manufactured geometry representation provided by CT scan measurements. 
		
		\begin{table}[H]
			\centering
			\begin{tabular}{|c|c|c|c|}\hline
				Specimen &  CAD-based porosity, [-] & Experimental porosity, [-] &CT-based porosity, [-] \\\hline
				$2\times32\times1$ & 0.756 & 0.638 $\pm$ 0.006 &  0.647\\\hline
				$2\times32\times2$ & 0.770 & 0.630 $\pm$ 0.004 &  0.639\\\hline
				$2\times32\times3$ & 0.775 & 0.677 $\pm$ 0.003 &  0.679\\\hline
				$2\times32\times4$ & 0.777 & 0.630 $\pm$ 0.002 &  0.671\\\hline
			\end{tabular}
		\caption{Porosity comparison of the beam specimens.}
		\label{tab:PorosityBeamsTogetherWithCADAndCT}
		\end{table}

		\paragraph{Direct numerical simulations of three-point bending test\newline}	 
		\renewcommand{\graphDir}{./sections/results/Pictures}
		In order to further support the above observations, we carry out a numerical simulation of the three-point bending test described in \cref{sec:experimentalTest}. Numerical experiments are performed for each one of the four specimens on both as-designed (CAD) and as-manufactured (CT) geometrical models. In both cases, the same boundary conditions and load cases are applied as in the experimental setup. The simulation of the as-designed geometry is carried out by using Comsol\texttrademark $\,$ with quadratic tetrahedral Finite Elements, whereas as-manufactured geometry is simulated using the high-order Finite Cell Method as described in~\cref{sec:NumericsFCM} with finite cells of polynomial degree $p=3$ containing $2\times2\times2$ voxels. A representative discretization is depicted in~\cref{fig:Mesh}. A total number of $51\times524\times32$ cells is used in this case.
			
		Representative displacement and von Mises stress distributions for an as-manufactured beam specimen are shown in~\cref{fig:BendingFields}. 
		
		\newcommand{\tikzDir}{./sections/results/Pictures/TikZ/Bending}
		\newcommand{\dataDir}{./sections/results/Pictures/TikZ/Bending}
		\begin{figure}[H]
			\centering
			\hspace*{-15mm}\subfloat[][Overview of the Finite Cell mesh with a representative displacement distribution ]{\includegraphics[scale=0.17]{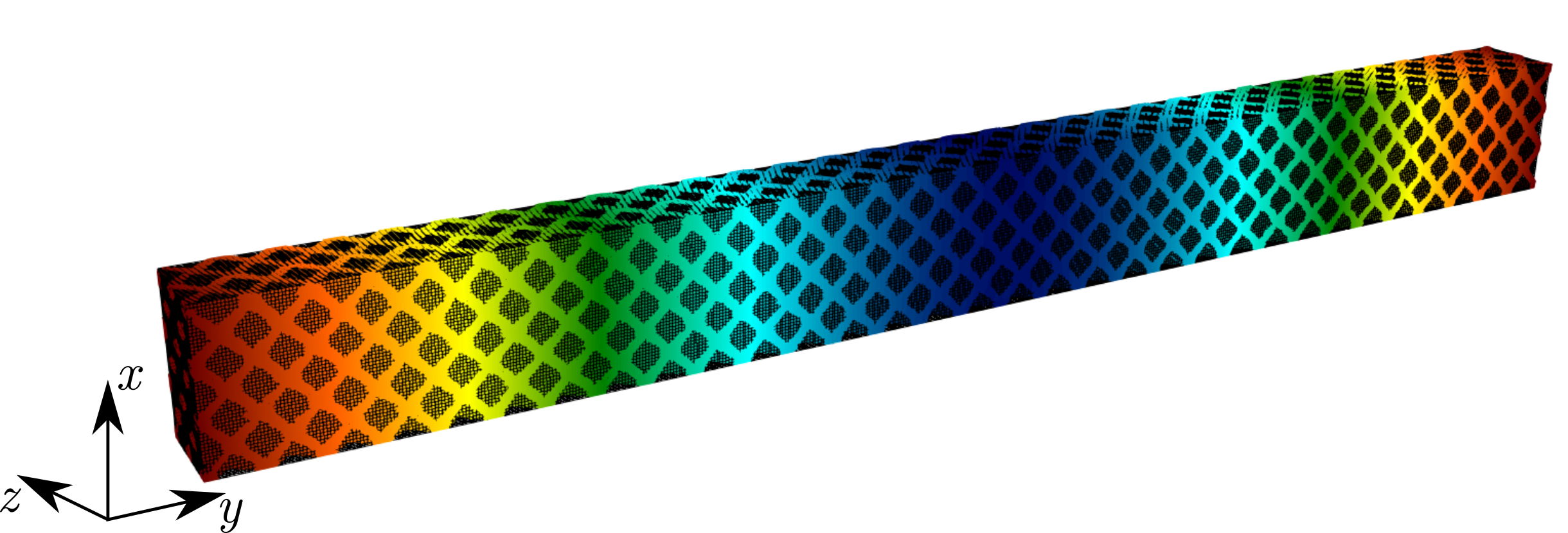}}\hspace*{5mm}
			\subfloat[][Zoom on the Finite Cells in the corner of the beam]{\includegraphics[scale=0.17]{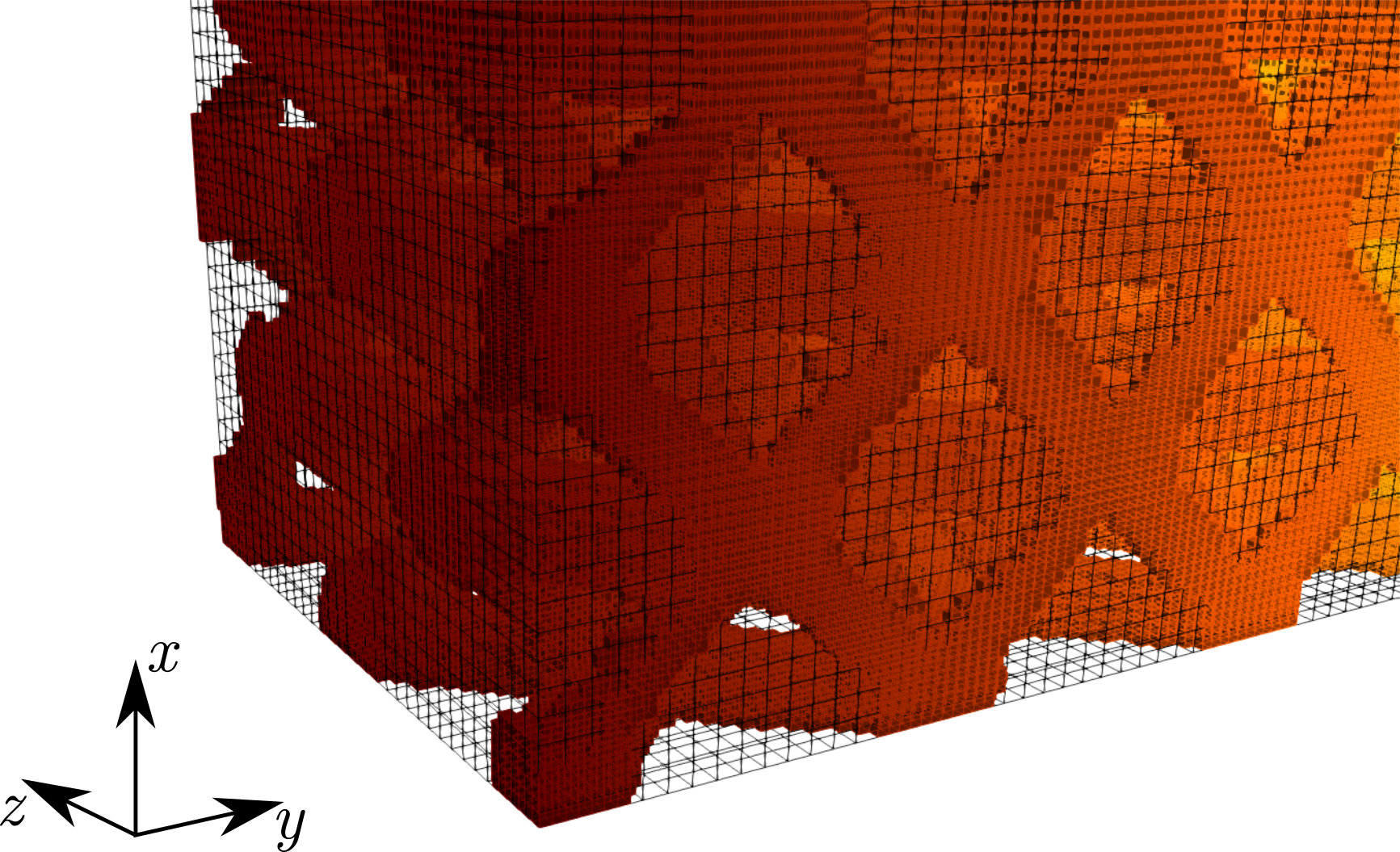}}\\
			\hspace*{-5mm}\subfloat[][Zoom on one periodic cell with Finite Cells (in black)]{\includegraphics[scale=0.23]{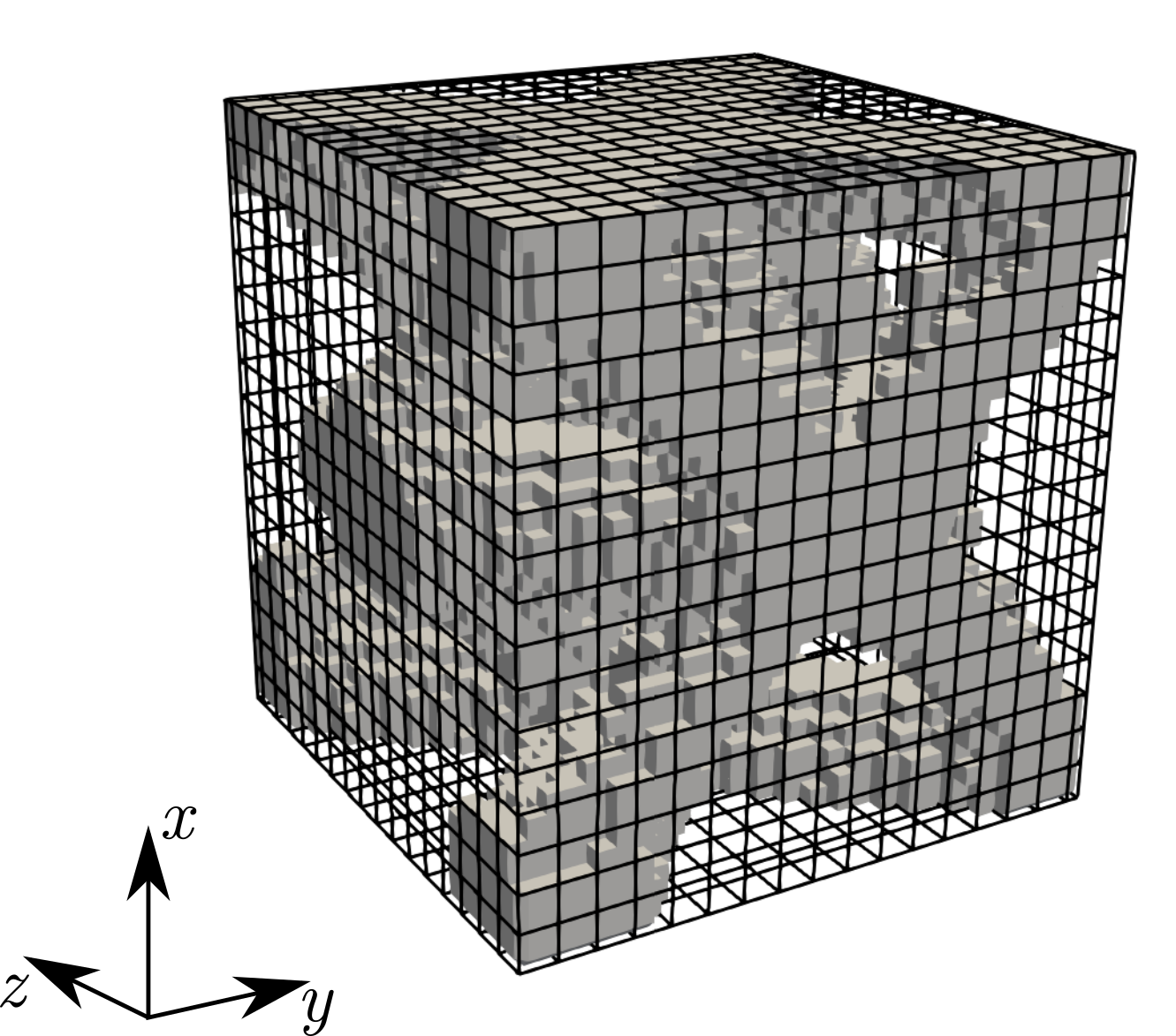}}\hspace*{25mm}
			\subfloat[Zoom on one periodic cell with Finite Cells (in black) and voxels (in blue)]{\includegraphics[scale=0.23]{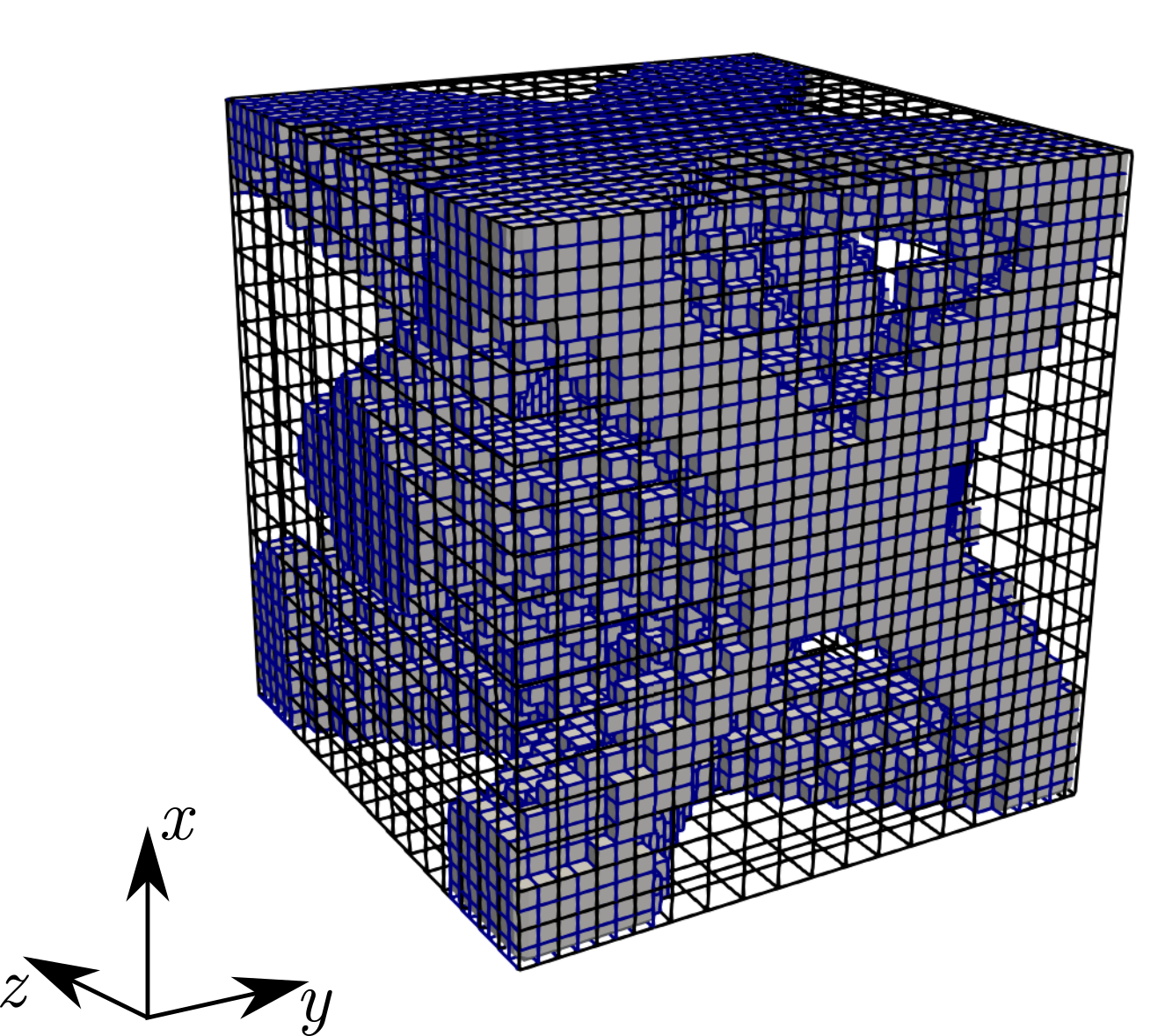}}
			\caption{Finite Cell mesh with $51\times524\times32$ cells for $2\times3\times32$ beam specimen.}
			\label{fig:Mesh}
		\end{figure}
		
		\begin{figure}[H]
			\centering
			\subfloat[][Displacement field in mm in the load direction]{\hspace*{5mm}\includegraphics[scale=0.15]{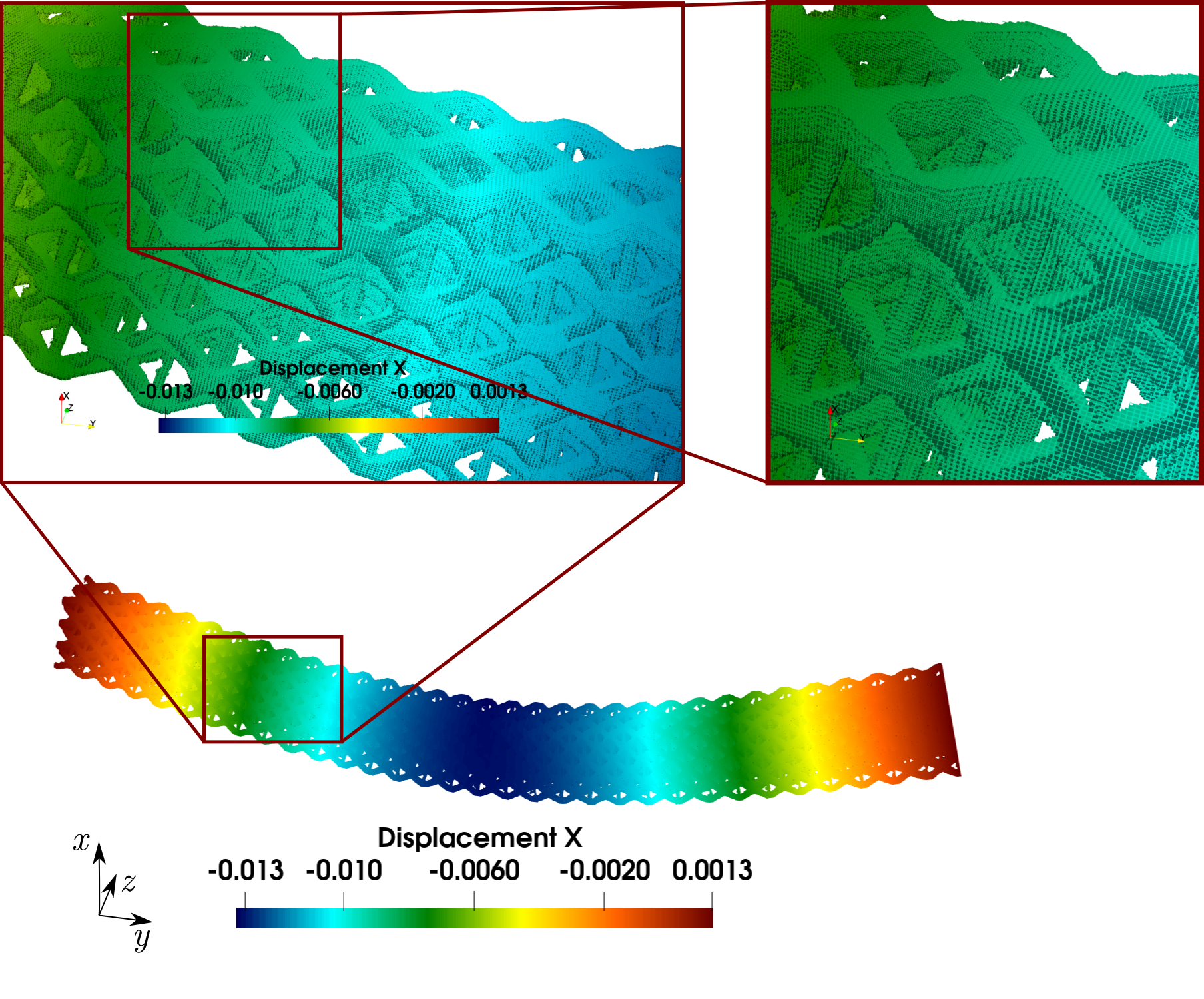}}
			\subfloat[][Von Mises stress in MPa Distribution]{\hspace*{8mm}\includegraphics[scale=0.161]{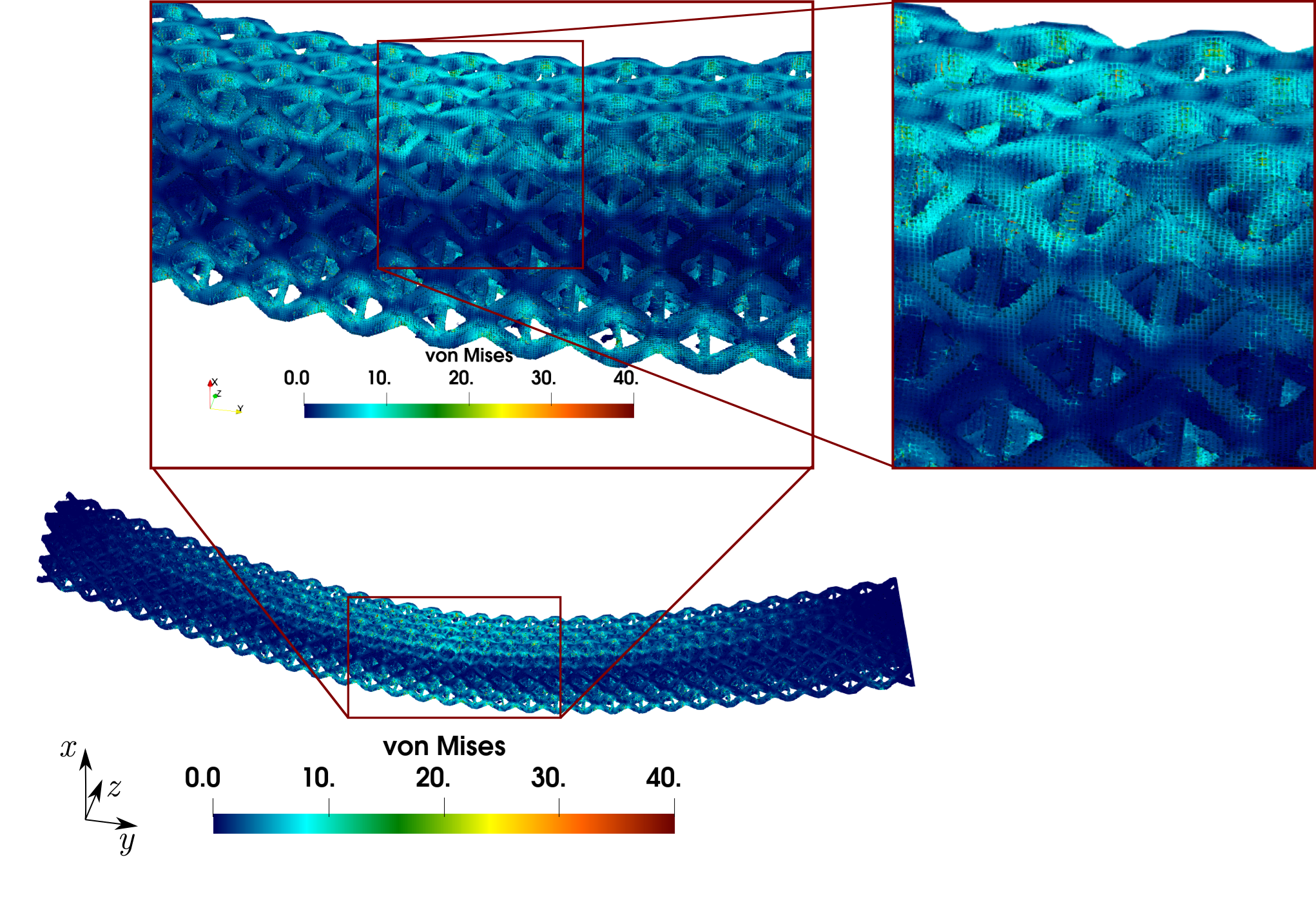}}
			\caption{Displacement and von Mises stress distributions for as-printed beam $2\times3\times32$ utilizing the Finite Cell Method.}
			\label{fig:BendingFields}
		\end{figure}
		
		The numerical bending rigidities are, then, computed by using~\cref{eq:bendingRigidity} and their values are compared to the experimental ones in~\cref{fig:BendingComparison}. 
		
		The qualitative comparison of these results shows that the as-designed and as-manufactured geometries follow the same tendency of a higher rigidity value for thicker beams. Nevertheless, quantitatively the relative errors in the bending rigidity value are always above $40\%$. This gap is largely driven by the geometrical difference between the as-manufactured and as-designed geometries. As the CT-based and experimental porosity values shown in~\cref{tab:PorosityBeamsTogetherWithCADAndCT} are lower than the designed ones, the as-designed bending rigidity should agree with this trend. According to the results in~\cref{fig:BendingComparison} the as-manufactured bending rigidity is larger than the designed one, thus, supporting the described tendency. Furthermore, the numerical simulation on the printed geometry via computed tomography provides an excellent agreement with the experimental tests, with a relative error always below 4\%.
		
		\begin{figure}[H]
			\centering
			\includegraphics[width=\textwidth,height=0.4\textheight]{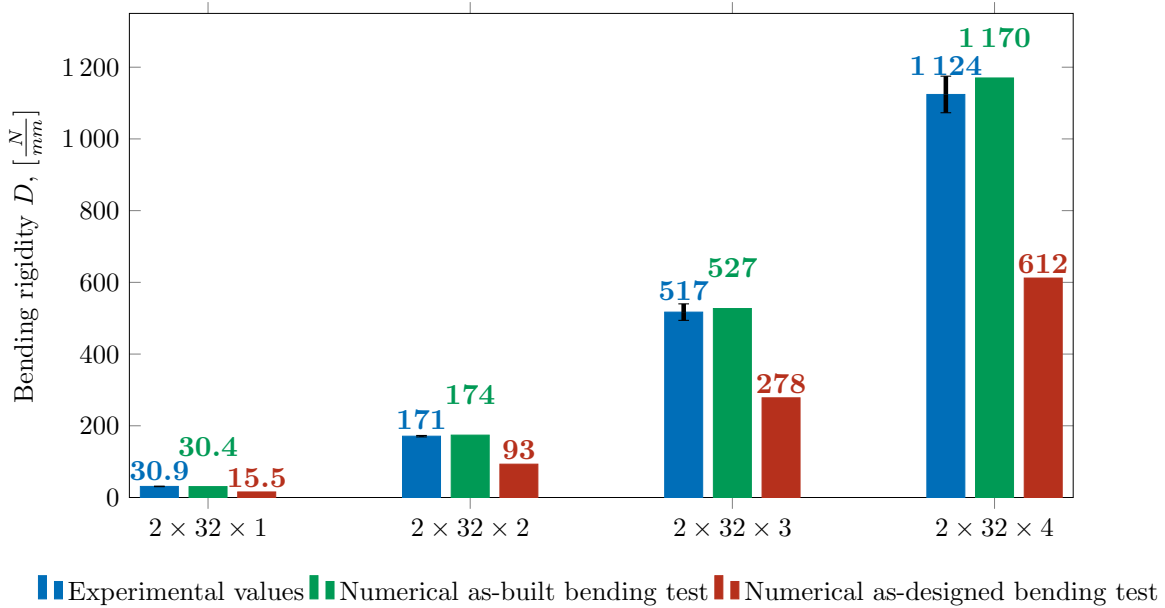}
			\caption{Comparison of bending rigidity obtained by numerical bending tests on the original as-designed geometry and on the as-manufactured geometry obtained from CT-scan data.}    
			\label{fig:BendingComparison}     
		\end{figure}
	}

\subsection{Experimental validation of strain gradient beam theory for octet-truss lattices}
\label{subsec:strainTheoryResults}
{
\newcommand{\tikzDir}{./sections/results/Pictures/TikZ/Bending}
\newcommand{\dataDir}{./sections/results/Pictures/TikZ/Bending}
Since in a three-point bending it is often desired to predict the mechanical behavior by dimensionally reduced beam models, we investigate more carefully the applicability of the beam models described in~\cref{sec:strainGradientTheory} to octet-truss lattice structures.

The beam models rely on the identification of effective quantities, such as Young's modulus $E\strut^{\ast}$ and shear modulus $G\strut^{\ast}$. As briefly mentioned in~\cref{sec:strainGradientTheory}, there are two ways to obtain the necessary quantities. For the as-designed geometries, only the first-order homogenization can be applied, as there is no possibility to perform experimental tests on it, while for the as-manufactured structures, the effective Young's and shear modulus can be measured experimentally. In the scope of this work, only the as-manufactured Young's modulus of octet-truss lattices is experimentally evaluated, whereas the effective as-manufactured shear modulus is determined by means of the first-order homogenization technique mentioned in~\cref{sec:strainGradientTheory}.~\Cref{tab::effectiveQuantities} summarizes the effective quantities used in the following. 

\begin{table}[H]
	\centering
	\begin{tabular}{|c|c|c|}
		\hline
		Effective quantity &  As-designed & As-manufactured \\\hline
		$E\strut^{\ast}$, MPa & $7\,356$  &  $12\,533\pm751$\footnote[2]{Experimental measure}  \\
		$G\strut^{\ast}$, MPa & $2\,742$ & $5\,651$  \\\hline
	\end{tabular}
	\caption{Effective mechanical quantities of the octet-truss specimens.\\ \footnotesize{{$ ^{\dagger} $}Experimental measure}}
	\label{tab::effectiveQuantities}
\end{table}

\Cref{fig:BendingGFitAll} shows the normalized bending rigidity $D/D^{EB}$ with respect to the beam height $h$ (see~\cref{eq:solutionrigidityEB}). The normalization is performed with respect to the Euler-Bernoulli bending rigidity ${D^{EB}}$ solution as follows:
\begin{equation}
\frac{D}{D^{EB}} = \frac{D w^{EB}}{F} = \frac{ w^{EB}}{w}
\end{equation}
where $w^{EB}$ is the classical Euler-Bernoulli solution for three-point bending as in~\cref{eq:solutiontEB}, $w$ is the experimentally recorded maximum deflection, and $D$ is the compared bending rigidity. 

As the as-manufactured and as-designed geometries have different effective properties, the bending rigidities are normalized with the Euler-Bernoulli solutions using the respective quantities from~\cref{tab::effectiveQuantities} and, thus, they are plotted separately in~\cref{fig:BendingGFit,fig:BendingGFitCAD}.

\begin{figure}[H]
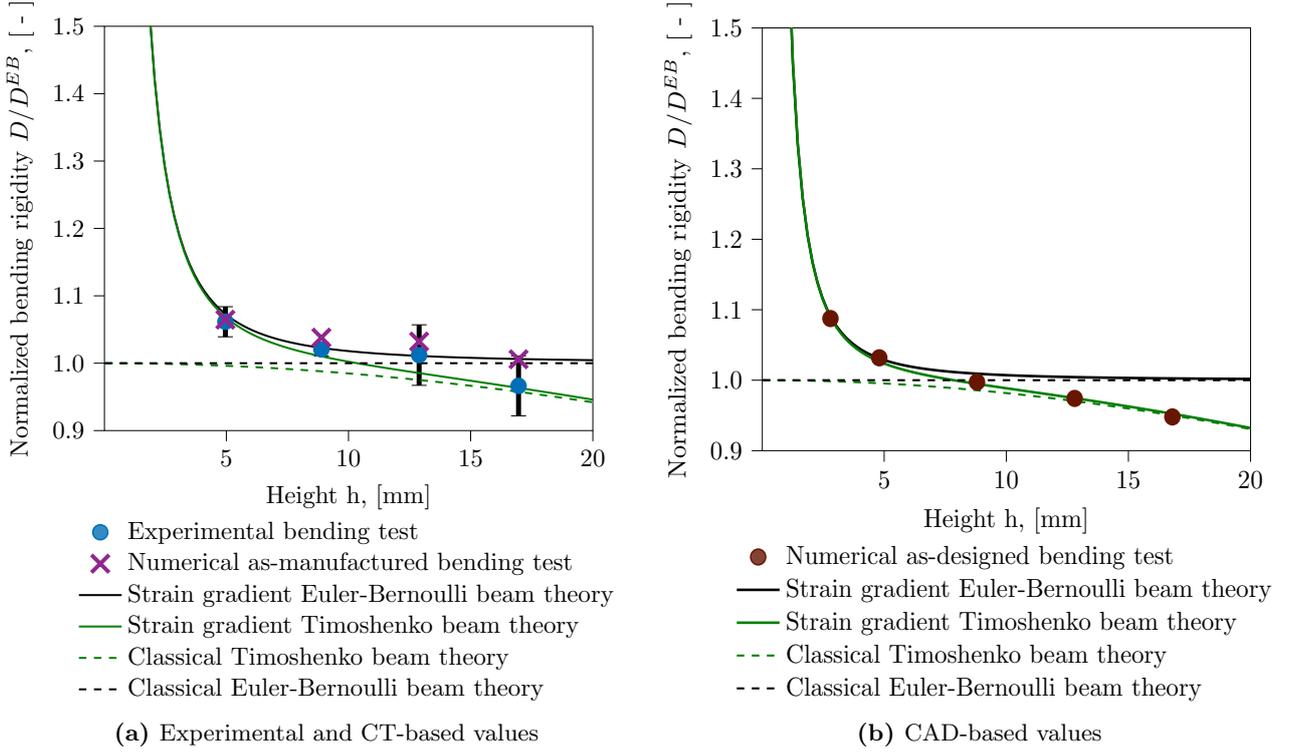

	\centering
	\subfloat[][Experimental and CT-based values]{\centering
		\resizebox{0.5\textwidth}{0.4\textheight}{%
			\includegraphics{\tikzDir/gFit.tikz}	
		}
		\label{fig:BendingGFit}  }
	\subfloat[][CAD-based values]{\centering
		\resizebox{0.5\textwidth}{0.4\textheight}{%
			\includegraphics{\tikzDir/Fitting_Results.tikz}	
		}	\label{fig:BendingGFitCAD}}
	\caption{Normalized bending rigidities of the octet-truss lattice beams with respect to the beam
		height.}
	\label{fig:BendingGFitAll}
\end{figure}

In both plots of~\cref{fig:BendingGFitAll}, the dashed lines indicate the results predicted by the classical beam theories, while the solid lines stand - the strain-gradient beam theories. The blue dots correspond to the experimental bending rigidity, whereas the crosses indicate the results of the numerical bending simulation computed on the as-manufactured specimen from~\cref{fig:BendingComparison}. Both values are normalized with the analytical Euler-Bernoulli solution using the as-manufactured effective Young's modulus from~\cref{tab::effectiveQuantities}. The brown dots in~\cref{fig:BendingGFitCAD} indicate the CAD-based results of the numerical bending test and again the results are normalized with the Euler-Bernoulli solution with the as-designed effective Young's modulus from~\cref{tab::effectiveQuantities}. Since as-designed geometry allows for further reduction of the considered thickness-to-length ratios, an extra point is added at the height of $2.4$ mm. This setup leads to a thickness-to-length ratio of $0.015$.

\paragraph{Classical beam theory using as-manufactured and as-designed geometry\newline}

As the normalization is performed with respect to the corresponding classical Euler-Bernoulli solution, the dashed black lines remain at the value $1$ for both as-manufactured and as-designed geometries. If the octet-truss lattice beams were to follow this behavior, all bending rigidities would lay on a straight line. However, neither as-manufactured nor as-designed values seem to comply with the assumptions of the Euler-Bernoulli theory. Thus, the classical Euler-Bernoulli theory cannot be applied to the characterization of the bending behavior of the considered octet-truss lattices.

The classical Timoshenko beam theory indicated with the green dashed line converges to the Euler-Bernoulli theory with the decreasing beam height. These states correspond to extremely slender beams, thus, making shear effects of minor importance. The as-manufactured geometry results as shown in~\cref{fig:BendingGFit} propose that only the thickest specimen with $2\times32\times4$ cells and the thickness-to-length ratio of $0.13$ follows the Timoshenko theory. However, the rest of the points do not follow this curve. The as-designed bending behavior as depicted in~\cref{fig:BendingGFitCAD} shows a similar trend, where for the thickest specimens the points lay on the curve. Although the Timoshenko beam theory seems to provide a better solution compared to Euler-Bernoulli, none of them can capture the observed bending behavior well.

\paragraph{Strain gradient beam theory using as-manufactured geometry\newline}

\Cref{fig:BendingGFit} indicates the presence of a stiffening effect. When the height of the beam is close to the characteristic size of the unit cell, the size effects affect the macroscopic bending behavior of the components and cause stiffer behavior in comparison to a standard prediction of the classical beam theories. This size-dependent bending phenomenon is precisely captured by the strain gradient beam theories on the as-manufactured geometries.

The strain gradient beam theories as described in~\cref{sec:strainGradientTheory} introduce an additional material parameter $g$. This high-order parameter is unknown a priori and can only be determined by a calibration of the solid lines to the obtained numerical and experimental solutions (or by other generalized homogenization procedures~\cite{Yang2020}). As mentioned in~\cite{Khakalo2018}, this intrinsic length parameter behaves as a material parameter and it is independent of loading, problem type, or the beam model. This quantity only depends on the underlying geometry. Thus, it must be the same for both strain gradient Timoshenko and Euler-Bernoulli theories. The value of the high-order material parameter $g$ is determined as $0.349$ [mm] for the as-manufactured octet-truss lattice (see~\cref{tab::GParameter}). This intrinsic length parameter characterizes the size effects in the octet-truss lattice structures via both Euler-Bernoulli and Timoshenko strain gradient beam theories. Its order is close to the smallest strut size diameter of the unit cell of $0.4$ [mm].

Although both strain gradient beam theories seem to capture an overall stiffening trend, it is important to know which theory is applicable. The numerical solution indicated with crosses seems to rather follow the Euler-Bernoulli approximation. However, the experimental data indicated with blue dots do not give a clear direction of which theory to follow. The first three points lay on the strain gradient Euler-Bernoulli theory, while the last point corresponding to the thickness-to-length ratio $0.13$ seems to be away from it. This can suggest that for the last configuration the strain-gradient Timoshenko theory is more appropriate. However, the measurement error bars on the experimental data indicate that both theories could be applicable for this setup and the last point can as well lay on the black solid line. Furthermore, the CT-based porosity value for the thickest beam is further away from the experimental one. Thus, it could lead to uncertainty in the computed bending rigidity. To further clarify this let us look at the as-designed results.

\begin{table}[H]
	\centering
	\begin{tabular}{|c|c|c|}
		\hline
		&  As-designed $g$, [mm]& As-manufactured $g$, [mm]\\\hline
		Octet-truss beam & 0.244  & 0.387  \\\hline
	\end{tabular}
	\caption{Comparison of as-designed and as-manufactured high-order intrinsic length parameter of the octet-truss specimen.}
\label{tab::GParameter}
\end{table}

%
%

\paragraph{Strain gradient beam theory using as-designed geometry\newline}

As already pointed out, the effective quantities obtained on the as-designed model are far from the experimentally determined bending rigidity and are depicted separately in~\cref{fig:BendingGFitCAD}.

Curiously, for the as-designed geometry, a weaker stiffening effect is observed. For the thickness-to-length ratio of $0.03$ (i.e., for the thinnest beam), the CAD-based results show about 8.4\% stiffening compared to the thickest observation, while the as-printed analysis indicated 9.5\%.

This is also reflected in the intrinsic high-order material parameter $g$. It is determined as $g=0.244$ mm in the same manner as for the as-manufactured geometries (see~\cref{tab::GParameter}). The most remarkable observation is that this high-order material parameter is lower than the one for as-manufactured geometries, similarly to the behavior already observed in the porosity values, the effective quantities, and the bending rigidity of the octet-truss specimens.

Furthermore, the as-designed numerical results seem to clearly follow the strain gradient Timoshenko theory, whereas the strain gradient Euler-Bernoulli curve does not provide an accurate solution to the overall bending behavior. Although it should be noticed that for the thickness-to-length ratios of two thinnest specimens ($h < 5$ [mm]) the strain gradient beam models are already very close to each other.

\paragraph{Comparison between as-manufactured and as-design results\newline}

All in all, the overall stiffening tendency is similar to the one observed from the experimental and as-manufactured numerical analysis. But the as-manufactured values are about 50\% higher than the designed ones as shown in~\cref{fig:BendingComparison}. The as-manufactured computations always lie within the uncertainty range of the experimental measurements, whereas as-designed numerical results never fall in this range. This rather large difference has been observed in similar studies conducted by the same authors on tensile behaviors of octet-truss lattices~\cite{Korshunova2020a}.

Moreover, when a closer study on the as-manufactured and as-designed geometries is undertaken, the stiffening trend differs. Firstly, we have observed that the considered octet-truss beams experience size effects, such that classical beam theories are not applicable to approximate the bending behavior, whereas strain gradient beam theories provide a much more accurate description. Secondly, the as-manufactured bending rigidities show a stronger stiffening effect than the designed ones, as also reflected in the intrinsic material parameter determined for both geometries. This observation well correlates to all other material characteristics determined by the authors. 
}
}

%% file: sections/conclusion/conclusion.tex
\section{Conclusions}
\label{sec:conclusions}
{
The numerical analysis of additively manufactured metamaterials can be prohibitively expensive and often impossible at full scale. In the present work, we have shown and validated an efficient numerical framework to incorporate complex as-manufactured geometries in a direct image-to-analysis workflow. The achieved numerical results are fully supported by the experimental tests performed on the octet-truss lattices. These findings suggest that in both direct numerical simulations and beam theories there is a strong need to incorporate as-manufactured geometries into the numerical analysis of AM products. In particular, the direct numerical simulation of CT-based as-manufactured geometries delivers results very close to the experimental measurements, whereas numerical analysis computed on the as-designed model fails to correctly predict the mechanical behavior of these metamaterials, presenting relative errors in the bending rigidity value always above $40\%$.

Furthermore, we have demonstrated the applicability of classical and strain gradient beam theories to the prediction of the bending behavior of AM octet-truss lattices. This work has confirmed that size effects arise in these metamaterials, thus raising the importance of the high-order continuum theories. Additionally, we validated the strain gradient beam theories in combination with the Finite Cell Method. In particular, a high-order intrinsic material parameter was determined directly from the numerical analysis of the as-manufactured geometries. As this material parameter is independent of the problem type, it can be used for the dimensionally reduced modeling of such octet-truss lattice components under different loadings and boundary conditions.

To conclude, the proposed numerical framework provides an accurate and flexible tool to analyze the behavior of as-manufactured metamaterials. Furthermore, these results represent an excellent initial step toward the validation of the strain gradient continuum theories in the field of additive manufacturing. In this line of research, we intend to incorporate the demonstrated technique into the analysis of the statistically similar CT models of such mechanical metamaterials in the future. This step would allow expanding the capabilities of the proposed image-to-material-characterization workflow.
	
}

%% file: template/acknowledgements.tex
We gratefully acknowledge the support of Deutsche Forschungsgemeinschaft (DFG) through the project 414265976 – TRR 277 C-01 and TUM International Graduate School of Science and Engineering (IGSSE), GSC 81. This work was partially supported by the Italian Minister of University and Research through the MIUR-PRIN projects "A BRIDGE TO THE FUTURE: Computational methods, innovative applications, experimental validations of new materials and technologies” (No. 2017L7X3CS) and "XFAST-SIMS" (No. 20173C478N). The authors would like to acknowledge the project "MADE4LO - Metal ADditivE for LOmbardy" (No. 240963) within the POR FESR 2014-2020 program. We also kindly acknowledge Eng. Alberto Cattenone and Prof. Stefania Marconi of the 3DMetal laboratory of the Department of Civil Engineering and Architecture of the University of Pavia for providing facilities for additive manufacturing and experimental testing (http://www-4.unipv.it/3d/laboratories/3dmetalunipv/). We further acknowledge Academy of Finland through the project Adaptive isogeometric methods for thin-walled structures (decision number 304122) as well as the August-Wilhelm Scheer Visiting Professors Program established by TUM International Center and funded by the German Excellence Initiative. The authors also gratefully acknowledge the Gauss Centre for Supercomputing e.V. (www.gauss-centre.eu) for funding this project by providing computing time on the Linux Cluster CoolMUC-2 and on the GCS Supercomputer SuperMUC-NG at Leibniz Supercomputing Centre (www.lrz.de). Finally, the authors gratefully acknowledge Giorgio Vattasso from LaborMet Due (http://www.labormetdue.it/) for his technical support in obtaining CT scan images.

%% file: 2020_BendingBehaviorOfAMLatticeStructures.bbl
\begin{thebibliography}{}

\bibitem[Abedian et~al., 2013]{Abedian2013}
Abedian, A., Parvizian, J., D{\"u}ster, A., Khademyzadeh, H., and Rank, E.
  (2013).
\newblock Performance of {{Different Integration Schemes}} in {{Facing
  Discontinuities}} in the {{Finite Cell Method}}.
\newblock {\em International Journal of Computational Methods}, 10(03):1350002.

\bibitem[{ASTM International}, 2017]{ASTME111}
{ASTM International} (2017).
\newblock {\em {ASTM E 111-17}: Standard Test Method for Young’s Modulus,
  Tangent Modulus, and Chord Modulus}.
\newblock West Conshohocken, PA: American Society for Testing and Materials.

\bibitem[Cao et~al., 2020]{Cao2020}
Cao, X., Jiang, Y., Zhao, T., Wang, P., Wang, Y., Chen, Z., Li, Y., Xiao, D.,
  and Fang, D. (2020).
\newblock Compression experiment and numerical evaluation on mechanical
  responses of the lattice structures with stochastic geometric defects
  originated from additive-manufacturing.
\newblock {\em Composites Part B: Engineering}, 194:108030.

\bibitem[Dallago et~al., 2019]{Dallago2019a}
Dallago, M., Winiarski, B., Zanini, F., Carmignato, S., and Benedetti, M.
  (2019).
\newblock On the effect of geometrical imperfections and defects on the fatigue
  strength of cellular lattice structures additively manufactured via selective
  laser melting.
\newblock {\em International Journal of Fatigue}, 124:348 -- 360.

\bibitem[Deshpande et~al., 2001]{Deshpande2001}
Deshpande, V., Fleck, N., and Ashby, M. (2001).
\newblock Effective properties of the octet-truss lattice material.
\newblock {\em Journal of the Mechanics and Physics of Solids}, 49(8):1747 --
  1769.

\bibitem[{du Plessis} et~al., 2020]{Duplesis2020}
{du Plessis}, A., Yadroitsava, I., and Yadroitsev, I. (2020).
\newblock Effects of defects on mechanical properties in metal additive
  manufacturing: A review focusing on x-ray tomography insights.
\newblock {\em Materials \& Design}, 187:108385.

\bibitem[D{\"u}ster et~al., 2017]{Duster2017}
D{\"u}ster, A., Rank, E., and Szab{\'o}, B.~A. (2017).
\newblock The p-version of the finite element method and finite cell methods.
\newblock In Stein, E., Borst, R., and Hughes, T. J.~R., editors, {\em
  Encyclopedia of {{Computational}} Mechanics}, volume~2, pages 1--35. {John
  Wiley \& Sons}, {Chichester, West Sussex}.

\bibitem[Echeta et~al., 2020]{Echeta2020}
Echeta, I., Dutton, B., Feng, X., Leach, R., and Piano, S. (2020).
\newblock Review of defects in lattice structures manufactured by powder bed
  fusion.
\newblock {\em International Journal of Advanced Manufacturing Technology},
  106:2649–2668.

\bibitem[Haubrich et~al., 2019]{Haubrich2019}
Haubrich, J., Gussone, J., Barriobero-Vila, P., K{\"u}rnsteiner, P., J{\"a}gle,
  E.~A., Raabe, D., Schell, N., and Requena, G. (2019).
\newblock The role of lattice defects, element partitioning and intrinsic heat
  effects on the microstructure in selective laser melted {Ti-6Al-4V}.
\newblock {\em Acta Materialia}, 167:136 -- 148.

\bibitem[Hua et~al., 2020]{Yang2020}
Hua, Y., Timofeev, D., Giorgio, I., and Mueller, W. (2020).
\newblock Effective strain gradient continuum model of metamaterials and size
  effects analysis.
\newblock {\em Continuum Mechanics and Thermodynamics}.

\bibitem[Jomo et~al., 2017]{Jomo2016}
Jomo, J., Zander, N., Elhaddad, M., {\"O}zcan, A.~I., Kollmannsberger, S.,
  Mundani, R.-P., and Rank, E. (2017).
\newblock Parallelization of the multi-level hp-adaptive finite cell method.
\newblock {\em Computers and Mathematics with Applications}, 74(1):126--142.

\bibitem[Jomo et~al., 2019]{Jomo2019}
Jomo, J.~N., {de Prenter}, F., Elhaddad, M., D'Angella, D., Verhoosel, C.~V.,
  Kollmannsberger, S., Kirschke, J.~S., N{\"u}bel, V., {van Brummelen}, E.~H.,
  and Rank, E. (2019).
\newblock Robust and parallel scalable iterative solutions for large-scale
  finite cell analyses.
\newblock {\em Finite Elements in Analysis and Design}, 163:14--30.

\bibitem[Khakalo et~al., 2018]{Khakalo2018}
Khakalo, S., Balobanov, V., and Niiranen, J. (2018).
\newblock Modelling size-dependent bending, buckling and vibrations of {2D}
  triangular lattices by strain gradient elasticity models: Applications to
  sandwich beams and auxetics.
\newblock {\em International Journal of Engineering Science}, 127:33 -- 52.

\bibitem[Khakalo and Niiranen, 2019]{Khakalo2019}
Khakalo, S. and Niiranen, J. (2019).
\newblock Lattice structures as thermoelastic strain gradient metamaterials:
  Evidence from full-field simulations and applications to functionally
  step-wise-graded beams.
\newblock {\em Composites Part B: Engineering}, 177:107224.

\bibitem[Korshunova et~al., 2020a]{Korshunova2020a}
Korshunova, N., Alaimo, G., Hosseini, S., Carraturo, M., Reali, A., Niiranen,
  J., Auricchio, F., Rank, E., and Kollmannsberger, S. (2020a).
\newblock A {CT}-based numerical characterization of tensile behavior of
  additively manufactured octet-truss structures and its experimental
  validation.
\newblock {\em Preprint submitted to Additive Manufacturing},
  https://arxiv.org/abs/2012.07452.

\bibitem[Korshunova et~al., 2020b]{Korshunova2020}
Korshunova, N., Jomo, J., L{\'e}k{\'o}, G., Reznik, D., Bal{\'a}zs, P., and
  Kollmannsberger, S. (2020b).
\newblock Image-based material characterization of complex microarchitectured
  additively manufactured structures.
\newblock {\em Computers \& Mathematics with Applications}, 80(11):2462 --
  2480.

\bibitem[Kudela et~al., 2016]{Kudela2016}
Kudela, L., Zander, N., Kollmannsberger, S., and Rank, E. (2016).
\newblock Smart octrees: {{Accurately}} integrating discontinuous functions in
  {{3D}}.
\newblock {\em Computer Methods in Applied Mechanics and Engineering},
  306:406--426.

\bibitem[Latture et~al., 2018]{Latture2018}
Latture, R.~M., Begley, M.~R., and Zok, F.~W. (2018).
\newblock Design and mechanical properties of elastically isotropic trusses.
\newblock {\em Journal of Materials Research}, 33(3):249–263.

\bibitem[Lazar and Po, 2015]{Lazar2015}
Lazar, M. and Po, G. (2015).
\newblock The non-singular green tensor of {Mindlin's} anisotropic gradient
  elasticity with separable weak non-locality.
\newblock {\em Physics Letters A}, 379(24):1538 -- 1543.

\bibitem[Lei et~al., 2019]{Lei2019}
Lei, H., Li, C., Meng, J., Zhou, H., Liu, Y., Zhang, X., Wang, P., and Fang, D.
  (2019).
\newblock Evaluation of compressive properties of {SLM}-fabricated multi-layer
  lattice structures by experimental test and $\mu$-{CT}-based finite element
  analysis.
\newblock {\em Materials \& Design}, 169:107685.

\bibitem[Liu et~al., 2017]{Liu2017}
Liu, L., Kamm, P., {Garc{\'i}a-Moreno}, F., Banhart, J., and Pasini, D. (2017).
\newblock Elastic and failure response of imperfect three-dimensional metallic
  lattices: The role of geometric defects induced by {{Selective Laser
  Melting}}.
\newblock {\em Journal of the Mechanics and Physics of Solids}, 107:160--184.

\bibitem[Lozanovski et~al., 2019]{Lozanovski2019}
Lozanovski, B., Leary, M., Tran, P., Shidid, D., Qian, M., Choong, P., and
  Brandt, M. (2019).
\newblock Computational modelling of strut defects in {SLM} manufactured
  lattice structures.
\newblock {\em Materials \& Design}, 171:107671.

\bibitem[Maconachie et~al., 2019]{Maconachie2019}
Maconachie, T., Leary, M., Lozanovski, B., Zhang, X., Qian, M., Faruque, O.,
  and Brandt, M. (2019).
\newblock {{SLM}} lattice structures: {{Properties}}, performance, applications
  and challenges.
\newblock {\em Materials \& Design}, 183:108137.

\bibitem[Mindlin and Eshel, 1968]{Mindlin1968}
Mindlin, R. and Eshel, N. (1968).
\newblock On first strain-gradient theories in linear elasticity.
\newblock {\em International Journal of Solids and Structures}, 4(1):109 --
  124.

\bibitem[Niiranen et~al., 2019]{Niiranen2019}
Niiranen, J., Balobanov, V., Kiendl, J., and Hosseini, S. (2019).
\newblock Variational formulations, model comparisons and numerical methods for
  {Euler}-{Bernoulli} micro- and nano-beam models.
\newblock {\em Mathematics and Mechanics of Solids}, 24(1):312--335.

\bibitem[O'Masta et~al., 2017]{Omasta2017}
O'Masta, M., Dong, L., St-Pierre, L., Wadley, H., and Deshpande, V. (2017).
\newblock The fracture toughness of octet-truss lattices.
\newblock {\em Journal of the Mechanics and Physics of Solids}, 98:271 -- 289.

\bibitem[Onck et~al., 2001]{Onck2016}
Onck, P., Andrews, E., and Gibson, L. (2001).
\newblock Size effects in ductile cellular solids. part {I}: modeling.
\newblock {\em International Journal of Mechanical Sciences}, 43(3):681 -- 699.

\bibitem[Parvizian et~al., 2007]{Parvizian2007}
Parvizian, J., D{\"u}ster, A., and Rank, E. (2007).
\newblock Finite cell method.
\newblock {\em Computational Mechanics}, 41(1):121--133.

\bibitem[Rashed et~al., 2016]{Rashed2016}
Rashed, M., Ashraf, M., Mines, R., and Hazell, P.~J. (2016).
\newblock Metallic microlattice materials: A current state of the art on
  manufacturing, mechanical properties and applications.
\newblock {\em Materials \& Design}, 95:518 -- 533.

\bibitem[Renishaw-PLC, 2020]{ManualReinshaw}
Renishaw-PLC (2020).
\newblock {\em Data sheet: SS 316L-0407 powder for additive manufacturing}.
\newblock Renishaw-PLC.

\bibitem[Schaedler and Carter, 2016]{Schaedler2016}
Schaedler, T.~A. and Carter, W.~B. (2016).
\newblock Architected cellular materials.
\newblock {\em Annual Review of Materials Research}, 46(1):187--210.

\bibitem[Sha et~al., 2018]{Sha2018}
Sha, Y., Jiani, L., Haoyu, C., Ritchie, R.~O., and Jun, X. (2018).
\newblock Design and strengthening mechanisms in hierarchical architected
  materials processed using additive manufacturing.
\newblock {\em International Journal of Mechanical Sciences}, 149:150 -- 163.

\bibitem[Tancogne-Dejean and Mohr, 2018]{Tancogne2018}
Tancogne-Dejean, T. and Mohr, D. (2018).
\newblock Elastically-isotropic truss lattice materials of reduced plastic
  anisotropy.
\newblock {\em International Journal of Solids and Structures}, 138:24 -- 39.

\bibitem[Vayssette et~al., 2019]{Vayssette2019}
Vayssette, B., Saintier, N., Brugger, C., {El May}, M., and Pessard, E. (2019).
\newblock Numerical modelling of surface roughness effect on the fatigue
  behavior of {Ti-6Al-4V} obtained by additive manufacturing.
\newblock {\em International Journal of Fatigue}, 123:180 -- 195.

\bibitem[Yan et~al., 2012]{Yan2012}
Yan, C., Hao, L., Hussein, A., and Raymont, D. (2012).
\newblock Evaluations of cellular lattice structures manufactured using
  selective laser melting.
\newblock {\em International Journal of Machine Tools and Manufacture}, 62:32
  -- 38.

\bibitem[Yang et~al., 2012]{Yang2012a}
Yang, Z., Ruess, M., Kollmannsberger, S., D{\"u}ster, A., and Rank, E. (2012).
\newblock An efficient integration technique for the voxel-based finite cell
  method.
\newblock {\em International Journal for Numerical Methods in Engineering},
  91(5):457--471.

\bibitem[Yoder et~al., 2018]{Yoder2018}
Yoder, M., Thompson, L., and Summers, J. (2018).
\newblock Size effects in lattice structures and a comparison to micropolar
  elasticity.
\newblock {\em International Journal of Solids and Structures}, 143:245 -- 261.

\bibitem[Zheng et~al., 2014]{Zheng2014}
Zheng, X., Lee, H., Weisgraber, T.~H., Shusteff, M., DeOtte, J., Duoss, E.~B.,
  Kuntz, J.~D., Biener, M.~M., Ge, Q., Jackson, J.~A., Kucheyev, S.~O., Fang,
  N.~X., and Spadaccini, C.~M. (2014).
\newblock Ultralight, ultrastiff mechanical metamaterials.
\newblock {\em Science}, 344(6190):1373--1377.

\end{thebibliography}
